 \documentclass[11pt]{article}

 \pdfoutput=1

 \usepackage[polutonikogreek,english]{babel}

 \usepackage{textgreek} 
 
\usepackage{bigints}

\usepackage[round]{natbib} 

\usepackage{graphicx,siunitx,xspace}
\usepackage{amsmath}
\usepackage{amssymb}
\usepackage{hyperref}
\usepackage{enumitem}
\usepackage{color}
\usepackage{enumitem}
\usepackage{float}
\usepackage{amsfonts,bm}
\usepackage{epstopdf}
\usepackage[mathscr]{euscript}

  \usepackage[normalem]{ulem}
  \usepackage{textcomp}

\usepackage[utf8]{inputenc}
\usepackage[T1]{fontenc}
\usepackage{pifont} 

\usepackage{chemformula}

\usepackage{float}
\usepackage{ifthen}
\begin{document}


\begin{center}
{\bf \Large
An English translation of the papers of Max 
\citet[][]{Planck_Heat_Temp_J_Phys_Theor_Appl_1911,
Planck_Energie_Temp_Phys_Z_1911}}
\\ \vspace*{2mm} \hspace*{-0mm}
{\bf \Large about 
``\,\underline{Energy and temperature}\,'' from the two papers}
\\ \vspace*{2mm} \hspace*{-0mm}
{\bf \Large
``\,\underline{\'Energie et temp\'erature}\,'' and }
{\bf \Large 
``\,\underline{Energie und Temperatur}\,''}
\\ \vspace*{2mm}
{\bf \Large
to provide a readable version of the French or German contents.
}
\\ \vspace*{4mm}
{\bf \large\color{black}
Translated by Dr. Hab. Pascal Marquet 
}
\\ \vspace*{2mm}
{\bf\bf\color{black}  \large Possible contact at: 
    pascalmarquet@yahoo.com}
    \vspace*{1mm}
    \\
{\bf\bf\color{black} 
    Web Google-sites:
    \url{https://sites.google.com/view/pascal-marquet}
    \\ ArXiv: 
    \url{https://arxiv.org/find/all/1/all:+AND+pascal+marquet/0/1/0/all/0/1}
    \\ Research-Gate:
    \url{https://www.researchgate.net/profile/Pascal-Marquet/research}
}
\\ \vspace*{-2mm}
\end{center}

\hspace*{65mm} Version-1 / \today

\vspace*{-4mm} 
\begin{center}
 --------------------------------------------------- 
\end{center}
\vspace*{-12mm}

\bibliographystyle{ametsoc2014}
\bibliography{Book_FAQ_Thetas_arXiv}

\vspace*{-3mm} 
\begin{center}
--------------------------------------------------- 
\end{center}
\vspace*{-3mm}

I have kept {the original text of Planck (in black)} unchanged, while sometimes including {\it\color{blue}{additional notes (in blue)}, including the (sub)sections} that were not included in the linear text of Max Planck.

The texts \dashuline{underlined with dashlines} were highlighted in German by Max Planck with separated letters (like: ``{\,... T\,e\,m\,p\,e\,r\,a\,t\,u\,r \:  u\,n\,d \:  E\,n\,e\,r\,g\,i\,e\, ...\,}'' instead of ``{\,... Temperatur und Energie ...\,''\,),
and in French with italic letters
(\, like \, ``{\it\,... temp\'erature et \'energie ...}\,''\,).

I have retained the symbols of the German paper 
like $d\left(\, W_1 \: W_2 \, \right) = 0$ and $1/T=dS/dU$, 
instead of something like 
$d\left( \, {\cal P}_1 \: {\cal P}_2 \, \right) = 0$ and
$1/\mathfrak{T}=dS/d{\cal E}$ written in the French paper (here with gothic+calligraphic letters, but even worse with French cursive letters). 

Do not hesitate to contact me in case of mistakes or any trouble in the English translation from the French or German texts (see the copy at the end of the document).


\vspace*{-3mm}
\begin{center}
\end{center}
\vspace*{-12mm}

  \tableofcontents

\vspace*{3mm}
\begin{center}
========================================================
\end{center}
\vspace*{-2mm}

\begin{center}
\underline{\Large\bf Energy and Temperature}
\\ \vspace*{2mm} {\Large\bf by Max Planck} 
\\ \vspace*{2mm}
{\large\bf (Lecture given on 21 April 1911
 at the French physics society.)}
\end{center}
\vspace*{-3mm}

\section{\underline{{\color{blue}Introduction (p.345)}}}
\label{Section-Intro}
\vspace*{-1mm}



Invited to present, in this place illustrated by a series of excellent scientific discoveries, one of the questions which have been the subject of my own work, I believe that I am doing my best to respond to the honour bestowed upon me, by exposing before you a problem which is precisely at the centre of thermodynamic research, the problem of \dashuline{the relationship between temperature and energy}. I have chosen this subject all the more willingly because it gives me the opportunity to pay my tribute of praise, in the very city where he was active, to the memory of the immortal Henri-Victor Regnault, to whom thermodynamics owes the determination of its most important numerical data, and whose centenary the world of physicists has just finished celebrating.


%
When we look at it from a distance, the question of the relationship between temperature and energy seems very simple and elementary; and it seems as easy to solve as it is to pose. I hope, however, to be able to show that this is not really the case. In reality, this problem conceals great difficulties and mysterious enigmas. Finally, quite recently, a new point of view has come to light, which will allow us to approach a little nearer the definitive solution of this fundamental problem of thermodynamics.
\vspace*{-2mm}

\section{\underline{{\color{blue}General thermodynamic arguments (p.345-347)}}}
\label{Section-Thermodynamic}
\vspace*{-1mm}


It has never been repeated too often that exact research in thermodynamics dates from the time when it was possible to distinguish the concepts of temperature and quantity of heat. 
The \dashuline{thermometer} and the \dashuline{calorimeter} have become the most important measuring instruments in thermodynamics. 
Both have received remarkable improvements over time, and with the precision of measurements has gone hand in hand the precision of definitions, without which the most delicate measurement would have no value. 
Indeed, while \dashuline{heat} was presented as a particular form of \dashuline{energy}, and knowledge of the mechanical equivalent of heat made it possible to express it in absolute quantities of energy, a definition of \dashuline{temperature} that was perfectly exact and suitable for precision measurements was deduced from the application to molecular heat, on the one hand, and to radiant heat, on the other, of the second principle of thermodynamics, the \dashuline{Carnot-Clausius} principle. 
We can therefore rightly admit that the quantities of heat, as well as the temperatures, can be determined with a precision, limited only by the quality of the instruments, the improvement of which is, moreover, constantly continuing.



But if it seems possible to resolve this first question with all the desirable approximation, we still have to ask ourselves what is the general relationship that exists between temperature and energy. By simply placing ourselves at the point of view of energetics, we could perhaps say: ``\,Temperature is a factor of energy. Temperature is to heat energy what force is to mechanical work and what potential is to electrical energy. The difference in temperature of two bodies indicates the direction of heat exchanges between these two bodies, exactly in the same way that mechanical force gives the direction of movement, or the difference in potential the direction of electric current.\,''



But in speaking in this way, we overlook an essential point. The movement can also occur in the opposite direction to the force, an electric current can flow in the opposite direction to the fall of potential, while a flow of heat energy occurring in the opposite direction to the fall of temperature is absolutely impossible. 
The existence of a quantity endowed with the properties possessed by temperature is already a unique fact in all of physics. The fact that two bodies, in calorific equilibrium with a third, are still in calorific equilibrium with each other, is not a fact that goes without saying, it is a very remarkable and very important circumstance. 
Indeed, nothing analogous is encountered in the case of electrical equilibria. We recognize it immediately when, taking a rod of copper and another of zinc, we plunge them by one end into diluted sulphuric acid, and then we join the other two ends by a metallic conductor. What is obtained is not a state of equilibrium, but an electric current, which lasts until the surface of the copper has changed noticeably.



It is known that all the particular relations expressing thermodynamic equilibria can be established, in a complete manner and absolutely in conformity with experience, by starting from the second principle of thermodynamics. But, in the problem which concerns us, we are not dealing only with this principle. Pure thermodynamics remains worthy of all our suffrages --it has enriched physical science with inestimable discoveries--, but it is no longer possible for us and we no longer have the right to stick to what it teaches us. 
For it informs us as little about the magnitudes and the relations between the magnitudes of the constants characteristic of the thermodynamic properties of bodies, as it does about the time necessary for the evolution of irreversible phenomena, such as conduction and heat radiation, diffusion and chemical reactions. 
According to pure thermodynamics, the ratio of the two specific heats could be $2$ as well as $100$, and the heat conductivity of a gas could be indifferently $100,000$ times greater or $100,000$ times smaller than that of a metal.
\vspace*{-2mm}

\section{\underline{{\color{blue}The Principle of Equipartition of Energy (p.347-349)}}}
\label{Section-Equipartition-Energy}
\vspace*{-1mm}


In this case, atomistic considerations are the only ones that can put us forward; and, in the first rank, comes the \dashuline{kinetic theory of gases}. 
This theory teaches us, as a consequence of the \dashuline{law of Gay-Lussac and Avogadro}, that the temperature of an ideal gas is represented by the average energy of the motion of each molecule, whatever the molecular weight.



This law, simple and easily accessible to intuition, already seems to shed some light on the big question we have posed of the relationship between energy and temperature. 
It is indeed obviously capable of being generalized for the cases of non-perfect gases, vapors, liquids and solids. 
It is enough to admit only that there is calorific equilibrium between two bodies when the particular molecules of the two bodies, placed one against the other, possess the same average energy of movement. 
It is easy to realize this, by representing that the molecules of the two bodies mutually shock each other and thus exchange their energies, so that finally a certain statistical equilibrium is established, in which the average energies of the movements have been equalized on both sides.



In fact, \dashuline{L. Boltzmann and J.-W. Gibbs}, using considerations relating to probabilities, managed to deduce from \dashuline{Hamilton}'s general equations a general law, which is today called \dashuline{the principle of equipartition of 
energy}{\color{blue}$\,$\footnote{{\color{blue}$\:$Named in the German paper as ``{\it\,Prinzip der gleichmässigen Energieverteilung}\,'' and thus ``{\it\,principle of uniform energy distribution\,''} in English. / P. Marquet.}}}.
It follows from this law that, in any statistical equilibrium relating to a system dependent on a large number of variables, each independent variable which influences the energy of the system must be attributed the same quantity of energy.



Now among the independent variables characteristic of the state of a solid, liquid or gaseous body, it is always necessary, according to the kinetic theory, to count the components of the velocity of its molecules. Consequently, in a system constituted by any number of solid, liquid or gaseous bodies and being in a state of statistical equilibrium, the average energy corresponding to each component of the velocity of each molecule is the same, according to the law of equipartition of energy, for the whole system. It is therefore evident that the thermodynamic condition of calorific equilibrium is in perfect agreement with the law of statistical mechanics, if the temperature of a body is taken, in a completely general way, as the measure of the average kinetic energy of any component of the velocity of any one of its molecules, or, to speak in a still more general way, as the measure of the average energy, corresponding to any one of the independent variables characteristic of the body.



In this way it seems that the law of equipartition of energy contains the definitive answer to the question of the relationship between energy and temperature; and one must admit that the extreme simplicity of this law, moreover quite intuitive, should easily lead physicists to attribute to it a fundamental role in thermodynamics. And this way of seeing is still authorized by various striking verifications that it has encountered.



If the temperature expresses at the same time the average energy  corresponding to a particular variable, we will obtain the total energy of a body by multiplying its temperature by the number of independent variables which determine its state. 
The heat capacity at constant volume is then given immediately by the number of its independent variables, or in other words by the number of degrees of freedom of the body in question. 
A still more particular circumstance: with the units ordinarily used for the quantity of heat, the temperature and the gram-molecule, the molecular heat still has a numerical value equal to the number of degrees of freedom of the real molecule.



In the case of a monatomic gas, for example, if we consider the atoms as material points, and if we disregard their mutual attractions, the only degrees of freedom which intervene in the expression of energy are the three components of the speed of the atom. 
Consequently, with the usual units, the atomic heat of such a gas considered under a constant volume is equal to $3$. 
In the case of a solid body, on the contrary, we must add, to the three components of the speed of an atom, the three coordinates which measure the displacement of the atom with respect to the equilibrium position and thus determine its potential energy. 
Consequently the atomic heat of a solid body is equal to $6$, which agrees fairly closely with the \dashuline{law of Dulong and Petit}. 
The deviations from this law, and in particular the increase in specific heat with temperature observed for all substances, would be explained by the introduction of new degrees of freedom, due to a greater relaxation of the bonds of the atom in the molecular group. 
The fact that this increase does not occur suddenly, but continues in a continuous manner, would be due to the fact that the molecular bonds do not relax all simultaneously, but only little by little.



After these remarkable successes, it is not surprising that \dashuline{Boltzmann} made the principle of equipartition of energy the centre of the kinetic theory of heat, and that, even today, many physicists incline towards the same view, and trust without fear to the future, for the solution of the particular difficulties and contradictions which still remain. It will be easy for me, however, to show that this point of view can no longer be maintained today, and that new experiments, joined to others older and known for a long time already, force us to withdraw from the law of equipartition of energy its role as a fundamental principle of statistical heat equilibrium.


\section{\underline{{\color{blue}Departures from Principle of Equipartition of Energy (p.349-352)}}}
\label{Section-Departures-Equipartition-Energy}
\vspace*{-1mm}

\subsection{\underline{{\color{blue}Departures: thermodynamics (p.349-351)}}}
\label{Subsection-Departures-thermodynamics}
\vspace*{-1mm}


I begin with the oldest difficulties. A biatomic molecule (such as that of oxygen, hydrogen or nitrogen) possesses, if we consider the atoms as perfectly free material points, $9$ degrees of freedom, namely: the $6$ components of the velocities of the two atoms and the $3$ projections of the distance of the atoms on the three coordinate axes. 
Now the molecular heat at constant volume is not $9$, as one would expect, but only $5$. 
And this is the case in all analogous cases. 
The molecular heat is always smaller than that which one would deduce from the number of degrees of freedom. 

This is not all, however. The molecule of a monatomic gas (mercury vapor for example) is certainly not a simple material point. 
To be convinced of this, it is enough to take a look at the spectrum of mercury with its innumerable fine lines. 
If each of these lines corresponded to only one degree of freedom, in the sense of the statistical theory of heat, the atomic heat at constant volume of mercury vapor should not be equal to $3$, as it is, but exceed $1,000$.



It is understandable that these difficulties did not escape \dashuline{Boltzmann}, and that he sought to account for the embarrassing multitude of these degrees of freedom, which so stubbornly refused to be demonstrated by calorific measurements. 
Now it was impossible to interpret the facts in such a way as to eliminate these degrees of freedom. 
He therefore tried to explain the extraordinary weakness of the influence that the movements of the atoms inside the molecule exert on the specific heat of this molecule, by a delay in the establishment of perfect statistical equilibrium. 
He supposed that, during the time necessary for the measurement of the specific heat, the vibrations of the elements of the molecule do not manage to produce significant modifications, and that calorific equilibrium is only established later and very slowly as the movement of the molecule continues, so that the phenomenon is no longer accessible to observation. 
According to this view, the temperature of a gas, absolutely protected from any heat exchange with the outside, should slowly change by itself. 
But no indication of such a phenomenon has ever been observed. 
On the contrary, the measurement of specific heats by means of rapid sound vibrations has given exactly the same results as direct calorimetric measurements.



Even more difficult is the situation in the case of solid bodies. 
In particular, those which are good conductors of heat and electricity: metals. 
According to the new theory of electrons, which has been verified in so many of its various consequences, it is admitted that the carriers of heat and electricity, evolving by conductivity, are electrons called free electrons, which can circulate between the molecules of the metal. 
If to such an electron one were to attribute, as is appropriate to an electron which really deserves the beautiful name of "free", the $3$ degrees of freedom corresponding to the three components of its speed, the molecular heat of a metal should always be appreciably greater than $6$.
\vspace*{-2mm}

\subsection{\underline{{\color{blue}Departures: the radiant heat (p.351-352)}}}
\label{Subsection-Departures-radiant-heat}
\vspace*{-1mm}


In all the examples given so far the contradiction remains in some way latent; and one could always hope that a happy modification of the principle would allow the difficulty to be removed once again. 
The first time that the law of equipartition of energy came into open conflict with experience was when it was attempted to apply it to the laws of black-body radiation. 
Radiant heat is not only measured more precisely than conduction heat, but also, thanks to spectral analysis, it can be studied in its most delicate characteristics, whereas conduction heat always presents itself as an indivisible whole. 
This is why it was reserved for research instituted in the field of radiant heat to bring a little more clarity into the study of the relationship between temperature and energy.



I believe that \dashuline{J.-H. Jeans} has demonstrated irrefutably that the law of equipartition of energy, when applied to the phenomena of heat radiation, leads to a distribution of energy in the normal spectrum characterised by the fact that the intensity in the spectrum is proportional to the temperature, and inversely proportional to the fourth power of the wavelength. 
Namely: the shorter the wavelength and the higher the frequency of vibration, the greater the number of degrees of freedom to which radiation distributed over a given spectral width corresponds.

We can see straight away that this energy distribution law does not lead to a truly determined energy distribution for the entire spectrum, and that it does not therefore correspond to the possibility of any true thermodynamic equilibrium. 
Indeed, for such an equilibrium to be achieved, the intensity of the spectrum would have to reach a maximum in every case, then decrease again and finally fade away as the wavelength decreases. 
Nevertheless, \dashuline{Jeans} did not conclude that the law of equipartition of energy could not be maintained; but, following the path that \dashuline{Boltzmann} had already indicated, he looked for a way out by considering the evolution in time of the phenomenon of radiation. 
According to him, when radiation takes place in an empty enclosure, a true thermodynamic equilibrium is not reached, but very short-wave radiation must constantly be formed, which is scattered outside, in the manner of Röntgen's hard rays. 
But this view is even more difficult to justify than \dashuline{Boltzmann}'s. 
Indeed, as \dashuline{O. Lummer} and \dashuline{E. Pringsheim}, it is in direct contradiction with all experiments. 
And, as all other avenues are closed, the generality of the law of equipartition of energy is undermined for the first time.

\subsection{\underline{{\color{blue}Departures: the specific heats (p.352-352)}}}
\label{Subsection-Departures-specific-heats}
\vspace*{-1mm}

This can be seen even more clearly. The specific heats of solid bodies show a persistent tendency to decrease with temperature; and, very recently, \dashuline{W. Nernst} has shown, either experimentally by measurements continued up to the boiling temperature of hydrogen, or theoretically by application of his new thermodynamic theorem: the specific heats of all solid and liquid bodies converge towards an infinitely small value, when the temperature decreases indefinitely. Thus the specific heat of copper at the boiling temperature of hydrogen is only one thirtieth of its value at ordinary temperature. How can we explain this fact using the principle of equipartition of energy? In particular, how can we explain the fact that the molecular heat of a body becomes smaller than 3? In fact, as soon as a molecule can move in a space, the components of its speed constitute 3 degrees of freedom of movement. The truth is that we could also suppose that, at low temperatures, the molecules of a solid or liquid body partly agglomerate into rigid groups, capable only of moving as a whole. This would reduce the number of degrees of freedom. But there are still all the movements from which the phenomena of emission and absorption of radiant heat of all wavelengths derive; and these are in any case very fluid, since they correspond to degrees of freedom, the number of which certainly exceeds the number of molecules by more than three.
\vspace*{-2mm}

\section{\underline{{\color{blue}Computation of probabilities (p.352-355)}}}
\label{Section-Computation-probabilities}
\vspace*{-1mm}



After all these facts have been set out, there can no longer be any doubt about the absolute necessity of this conclusion: the law of equipartition of energy does not play the fundamental role in thermodynamics that it has been attributed for some time. The question of the relationship between temperature and energy is thus once again posed with all its acuteness. If, in statistical equilibrium, the average energy is not \dashuline{uniformly}  distributed between the different degrees of freedom, the average energy corresponding to a particular degree of freedom cannot be taken as a measure of temperature, since in the state of equilibrium the temperature must certainly be the same everywhere.


How can we now get out of this difficult dilemma? Should we regard the law of equipartition of energy as completely false, and seek something entirely new? No, certainly not. This law has indeed received in a certain domain, in particular in the case of monatomic gases and even to a certain extent in the case of solid bodies, striking verifications. It therefore certainly contains a part of truth. But it does not contain the whole truth. We are thus led to think that it constitutes an illegitimate generalization of a principle that is correct in itself. And, in order to arrive at the exact generalization, we shall have to follow again the path that led to the law of equipartition of energy, find the point where we embarked on the wrong path, take care not to go astray there, and take the right direction. In order to put this plan into practice, we will first note that the law of equipartition of energy has been derived from the application of the \dashuline{calculus of probabilities} to statistical equilibrium. We will also retain this starting point, because without the introduction of statistical considerations it is impossible to grasp the characteristic particularities of calorific equilibrium, contrary to what has taken place for mechanical or electrical equilibrium.


The state of statistical equilibrium is distinguished from all other possible states, corresponding to the same sum of energy, in that it is conditioned by a maximum probability. If two bodies, isolated from all the rest, can mutually exchange their heat energies by conductivity or by radiation, they will be in statistical equilibrium with each other when the passage of heat from one body to the other no longer corresponds to an increase in probability. If $W_1=f(U_1)$ represents the probability that the first body has the energy $U_1$, $W_2=\varphi(U_2)$ the probability that the second body has the energy $U_2$, the probability that the first body has the energy $U_1$ and that at the same time the second body has the energy $U_2$ will be represented by $W_1\:.\:W_2$. The condition for this quantity to pass through a maximum gives: 
$$ d\left( W_1\:W_2 \right) \;=\; 0 
\;\;\;\;\;\;\mbox{or}\;\;\;\;\;\;  
  \frac{d\,W_1}{W_1} \;+\; \frac{d\,W_2}{W_2} \;=\;0 \; .$$
%
%
%
%
By adding the permanent condition: 
$$ d\,U_1 \;+\; d\,U_2 \;=\; 0 \; ,$$ 
which expresses that the total energy $U_1\,+\,U_2$ does not change, we obtain the equation 
$$ \frac{1}{W_1}\frac{d\,W_1}{d\,U_1} \;=\; 
   \frac{1}{W_2}\frac{d\,W_2}{d\,U_2}  $$
as a condition of statistical equilibrium.



If now we identify statistical equilibrium with calorific equilibrium, and if we note that the condition of calorific equilibrium is expressed by the equality of the temperature of the two bodies, we see immediately that we will obtain a perfect agreement between statistical theory and thermodynamics by taking, in a general way, the quantity 
$$ \frac{1}{W} \; \frac{d\,W}{d\,U} \;=\; \frac{d\ln(W)}{d\,U} \;$$ 
as a \dashuline{measure of the temperature} of a body, and by posing consequently: 
\begin{align}
k\; \frac{d\ln(W)}{d\,U} \;=\; \frac{1}{T} \; .
\label{Eq_1}
\end{align}



The fact of taking precisely the inverse, and not another function of temperature, comes only from the adoption of the ordinary temperature scale. In principle, any other temperature function would provide us with the same services. The constant $k$ depends only on the arbitrarily chosen units for temperature and energy.



We may regard equation (\ref{Eq_1}) as \dashuline{the most general answer} to the question we have posed about the relation between temperature and energy. It is evidently very closely related to the well-known equation of thermodynamics 
    $$ \frac{1}{T}  \;=\; \frac{d\,S}{d\,U} \; ,$$ 
however, it has quite a different meaning. The equation of pure thermodynamics in fact serves only as a definition of the entropy $S$ so that, taken in itself, it does not represent a physical law. In equation (\ref{Eq_1}), on the contrary, we have a real relation between the quantities defined independently of each other. We thus arrive by this means at a definition of entropy different from that of pure thermodynamics: 
    $$ S \;=\; k \: \ln(W) \; , $$ 
and this definition gives the physical meaning of entropy a much more intuitive one than the purely thermodynamic definition, the true meaning of which is grasped with great difficulty by so many young students. But, for the purpose we are pursuing here, we can stick to the consideration of the 
probability{\color{blue}$\,$\footnote{$\:${\color{blue}badly written $W_1$ in the French version / P. Marquet}.}}
$W$ alone and leave completely aside the auxiliary  
quantity{\color{blue}$\,$\footnote{$\:${\color{blue} i.e. the entropy / P. Marquet}.}} $S$.
\vspace*{-2mm}

\section{\underline{{\color{blue}Examples of computation of probabilities (p.355-358)}}}
\label{Section-Examples-computation-probabilities}
\vspace*{-1mm}


It seems that, thanks to the new relation obtained, our problem has made fundamental progress. It could even be considered as completely solved, if we could really give ourselves the probability $W$ as a function of the energy $U$. 
But this is not yet possible in a general way. 
There are, however, a certain number of important cases in which we can calculate the probability $W$ to the end. The methods which lead to such calculations were first developed by \dashuline{Boltzmann} and \dashuline{Gibbs} by admitting the legitimacy of the application of \dashuline{Hamilton}'s general equations and of the theorem which was deduced from them by \dashuline{Liouville}.
\vspace*{-2mm}

\subsection{\underline{{\color{blue}Examples of computation of probabilities: perfect gases (p.355-356)}}}
\label{Section-Examples-computation-probabilities-pefect-gases}
\vspace*{-1mm}


I first consider the case of an ideal gas, consisting of $N$ monatomic molecules. We then have 
$$ W \;=\; U^{\,3\,N/2} \:.\; const \; , $$
where the constant does not depend on the energy $U$, and substituting in equation (\ref{Eq_1}) gives:
$$ \frac{1}{T} \;=\; k \:.\: \frac{1}{U} \:.\: \frac{3\:N}{2}
 \;,\;\;\;\; U \;=\; k \:.\: \frac{3\:N}{2} \:.\: T \; .$$



If $k$ is known, we can calculate the absolute number $N$ of molecules by directly measuring $U$ and $T$. The average energy of an atom is then $(3\,k/2)\,T$ and the atomic heat $3\,k/2$. It is therefore independent of the nature of the gas, and corresponds perfectly to what is given by the law of equipartition of energy for three degrees of freedom, since each degree of freedom has the heat capacity $k/2$.
\vspace*{-2mm}

\subsection{\underline{{\color{blue}Examples of computation of probabilities: polyatomic gases (p.356-356)}}}
\label{Section-Examples-computation-probabilities-poly-gases}
\vspace*{-1mm}


If we wish to calculate the probability $W$ for a polyatomic gas, it is necessary to make an assumption about the number of degrees of freedom with which atoms, ions and electrons move in the molecule. It is here that we have clearly taken the wrong path, arriving at the law of equipartition of energy. If, in fact, we admit that the constituent particles of the molecule move as perfectly free points, and we apply \dashuline{Hamilton}'s equations to them, we obtain for the probability $W$ an expression of absolutely the same form as in the case of a monatomic gas. Only, instead of the number 3, we have the number $n$ of degrees of freedom of the molecule, and we arrive naturally at the law of equipartition of energy, that is to say, at a manifest contradiction with experience.

This is where the correction must be made. However natural and seductive may be the hypothesis, constantly made up to now, that \dashuline{Hamilton}'s equations still apply with all rigour to the delicate phenomena which take place inside molecules and even atoms, we must conclude that it constitutes an extrapolation condemned by experience, and suppose on the contrary that the number of degrees of freedom in the molecule, used to determine the probability $W$, is smaller, and often much smaller, than the number of elements in the molecule. 
This is only possible on condition that we take a completely different view of intramolecular phenomena from those we have used up to now. We then have to imagine a new hypothesis, one that has the effect of significantly limiting the number of different possible states within the molecule. 
The justification of such a hypothesis, by virtue of its absolute novelty, can only be made \dashuline{a posteriori}, and depends exclusively on experiment. 
Moreover, any hypothesis is permitted that does not contradict the known laws of physics. 
And, as we still know very little about intramolecular processes, there is still a vast field for the imagination.
\vspace*{-2mm}

\subsection{\underline{{\color{blue}Examples of computation of probabilities: molecules/radiation (p.356-358)}}}
\label{Section-Examples-computation-probabilities-radiation}
\vspace*{-1mm}



We shall have a hypothesis which reduces the number of degrees of freedom within a molecule, by supposing that the rapid vibrations, which originate within the molecule and produce the phenomena of emission and absorption of heat, cannot possess any energy, but that their energy is necessarily an \dashuline{integer multiple} of a certain finite quantity $\varepsilon$, which is determined by the frequencies of vibration. 
This hypothesis gives the expression 
$$ W \;=\; 
\frac{\displaystyle \left(\frac{U}{N\:\varepsilon}\:+\:1\right)^{U/\varepsilon+N}}
{\displaystyle \left(\frac{U}{N\:\varepsilon}\right)^{U/\varepsilon}} $$ 
for the probability of $N$ molecules possessing the energy $U$, and, according to equation (\ref{Eq_1}), between the temperature $T$ and the 
energy{\color{blue}$\,$\footnote{{\color{blue}$\:$This notation $U$ in (\ref{Eq_1}) in the German paper was previously wrongly written as $\varepsilon$ in the French paper. / P. Marquet.}}}
$U$, we have 
$$ U \;=\; \frac{N\:\varepsilon}{\displaystyle \exp\left(\frac{\varepsilon}{k\:T}\right) \;-\; 1} \;.$$.


This formula allows us to calculate directly the intensity of monochromatic radiation of the corresponding wavelength. We therefore have a means of verifying it by comparing the result of the calculation with the experimental laws of the distribution of energy in the normal spectrum of heat radiation. Experience has so far given perfect agreement, if we make $\varepsilon$ proportional to the vibration frequency $\nu\,$: 
$$ \varepsilon \;=\; h \; \nu \;\; .$$
%
%
We then 
have:{\color{blue}$\,$\footnote{{\color{blue}$\:$These 1911 values were already accurate up to $1.1$~\% and $2.5$~\%, respectively, when compared with the (exact) 2019 modern values in SI units:
$h=6.626\,070\,15 \times 10^{-34}$~J\,.\,s \;and\;
$k_B=1.380\,649 \times 10^{-23}$~J\,.\,K${}^{-1}$
(with the old cm-g-s unit $1$~erg$\:=10^{-7}$~J). / P. Marquet.}}}
$$ h \;=\; 6.55\:.\:10^{-27}\mbox{~erg-sec}
     \;\;\;\;\;\;\;\mbox{and}\;\;\;\;\;\;\; 
   k \;=\; 1.346\:.\:10^{-16}\mbox{~erg/deg} \; .$$

We can see that there is no longer any question of an equal distribution of energy. 
In fact, for two molecules with different vibration frequencies $\nu$, the average energies are quite different for the same temperature. 
But when the elementary quantity of energy $\varepsilon$ is very small, i.e. in the case of very slow vibrations, or when the temperature $T$ is very high, we obtain 
$$ U \;=\; N \: k \: T \; ,$$ 
and we arrive at the same relationship between temperature and energy as that which is deduced from the law of equipartition of energy, by admitting two degrees of freedom (for kinetic energy and potential energy). In general, the law of equipartition of energy does not apply exactly, and temperature cannot measure the average energy of a molecule.


\dashuline{A. Einstein} further hypothesized that, in the case of solid bodies, the vibration energy $U$ of the molecules, multiplied by the numerical factor $3$, corresponding to the three possible directions of vibration in space, then constitutes the total heat energy of the body. \dashuline{W. Nernst} studied experimentally, by measurements carried out at low temperatures, the characteristic formula for specific heats which is deduced from this hypothesis and verified it in its essential 
aspects{\color{blue}$\,$\footnote{\label{footnote-Einstein-Cp}{\color{blue}$\:$Note that the decrease suggested by A. Einstein (1907) for the specific heat capacity $c_v(T) \propto y^2\:\exp(-y)$ (with low values of $T$ and large values of $y=\beta\,\nu/T$) was only approximately verified by experimental values (the exponential decrease was too rapid for small $T$). 
The same will be true for the next Nernst-Lindemann (1911) formula (with half the sum of two Einstein's function). 
The true relationship for $c_v(T)$ will be derived next by Debye (1912), as an improvements of the previous Einstein and Nernst-Lindemann methods (a weighted integral over a limited range of frequencies of Einstein's functions), with as a result a less rapid decrease for $c_v(T) \propto T^3$ for $T$ approaching $0$~K. / P. Marquet.}}}.
The hypothesis of the elements of energy thus received a solid and completely new confirmation.
\vspace*{0mm}

\section{\underline{{\color{blue}Conclusions (p.358-359)}}}
\label{Section-Conclusions}
\vspace*{-1mm}


Although it has so far been verified by experiment, this hypothesis is still in its present form susceptible of 
improvement{\color{blue}$\,$\footnote{{\color{blue}$\:$See the footnote~\ref{footnote-Einstein-Cp}. / P. Marquet.}}}.
Indeed, the assumption that the vibration energy $U$ is an integer multiple of $\varepsilon$, leads to the following consequence. The molecule can change its vibration energy only by discontinuous jumps. It is then very difficult if not impossible to understand how the molecule can absorb at one go the entire elementary quantity of energy $\varepsilon$, while however, for the absorption of a finite quantity of energy, coming from a radiation of finite intensity, a finite duration is always necessary.



It therefore seems to me necessary to modify the hypothesis of the energy elements in the following way.
Only the \dashuline{emission} of energy takes place in bursts, in whole quantities of energy $\varepsilon$, and according to the laws of chance; absorption, on the other hand, continues in a perfectly continuous manner.

It is sometimes said in favour of a hypothesis that it is useful not only in the cases for which it was made, but also in other questions. Well, we can provisionally give this good mark to the hypothesis of energy elements. It is assumed here that a molecule can only emit vibrational energy in certain given quantities $\varepsilon$, whether it be pure radiation energy as in calorific radiation, Röntgen rays and $\gamma$ rays, or corpuscular radiation, as in the case of cathode rays and $\alpha$ and $\beta$ rays. 
Now, not only has this hypothesis already been verified with regard to the law of calorific radiation, but it has also provided a very precise method for determining the elementary quantities of electricity and matter, and has given us the key to penetrating the meaning of Nernst's thermodynamic 
theorem{\color{blue}$\,$\footnote{{\color{blue}$\:$Here Max Planck referred to the ``\,heat theorem\,'' of Walter Nernst (1906), which stated that: both the derivatives $d(\Delta S)/dT$ and $d(\Delta Q)/dT$ are zero in the vicinity of $0$~K for all chemical reactions, namely that both the change in entropy $\Delta S$ and the heat of reaction $\Delta Q$ have horizontal tangents at $0$~K when they are plotted against the absolute temperature $T$. 
Max Planck also referred to the next extension he will suggest (after 1911) and is nowadays known as the ``\,third law of thermodynamics\,'' (i.e. that the entropy $S$ itself tends toward zero for the more stable condensed state for $T$ becoming close to $0$~K). / P. Marquet.}}}.
It also seems to have played a fundamental role in the emission of cathode rays, in the photo-electric effect, and in the phenomena of radioactivity, whose strangeness borders on the prodigious, and to which the names of \dashuline{Becquerel, Currie, Rutherford} will always be attached.



Are we then to say that the hypothesis of elementary quantities of energy really contains the whole truth? Such an assertion would be audacious as well as indicative of a narrow-minded view. Nevertheless, I believe that this hypothesis is closer to the truth than the law of equipartition of energy, which, in its light, appears only as one of its particular cases. This is all that can reasonably be required of a new hypothesis. As for the definitive judgement on the value of this hypothesis, it is here, as in all questions of physics, up to experiment to make it.

\vspace*{-3mm} 
\begin{center}
--------------------------------------------------- 
\end{center}
\vspace*{-3mm}

%

%


\newpage
\section{\underline{\color{blue} The French Paper of Planck (1911)}}
\label{Section-French-PDF-1911}
\vspace*{0mm}

\begin{figure}[hbt]
\centering
\includegraphics[width=0.63\linewidth]{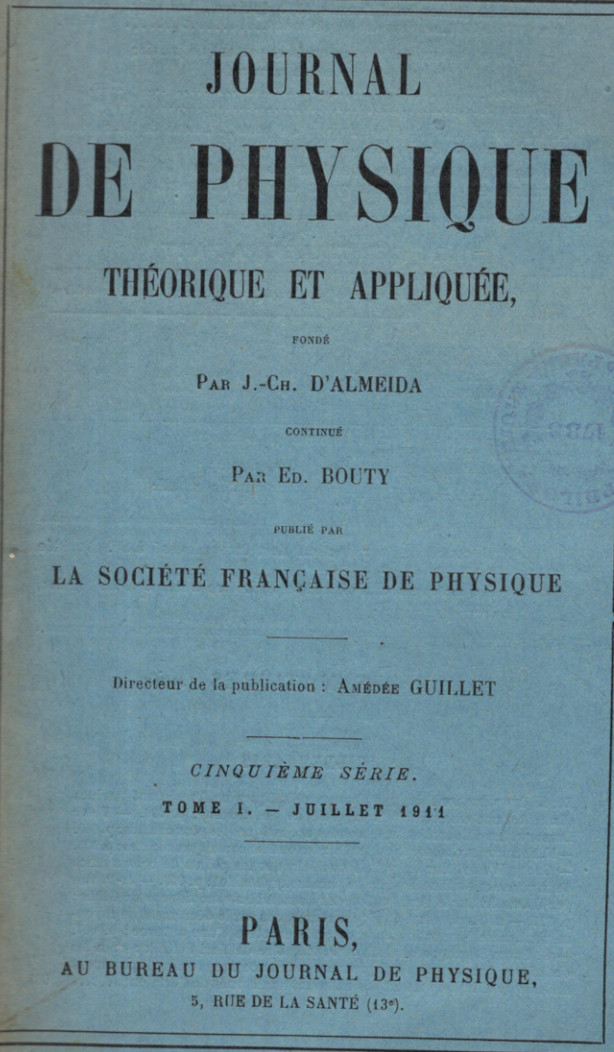}
\end{figure}
\begin{figure}[hbt]
\centering
\includegraphics[width=0.77\linewidth]{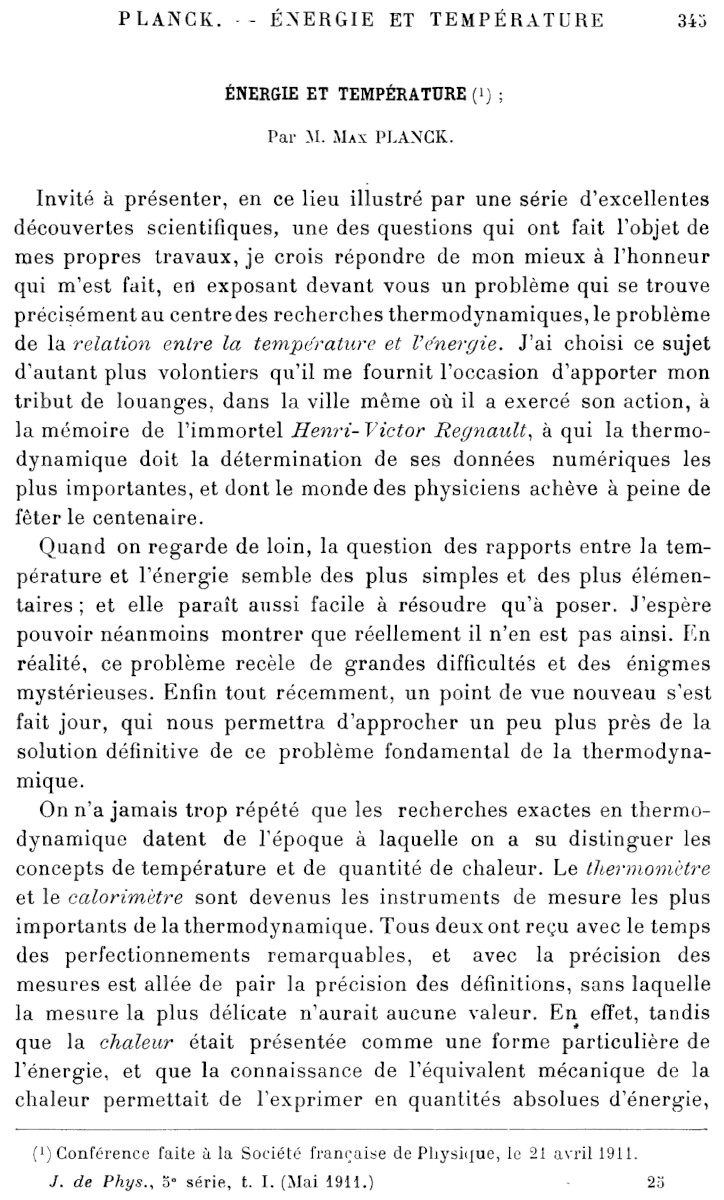}
\end{figure}
\begin{figure}[hbt]
\centering
\includegraphics[width=0.77\linewidth]{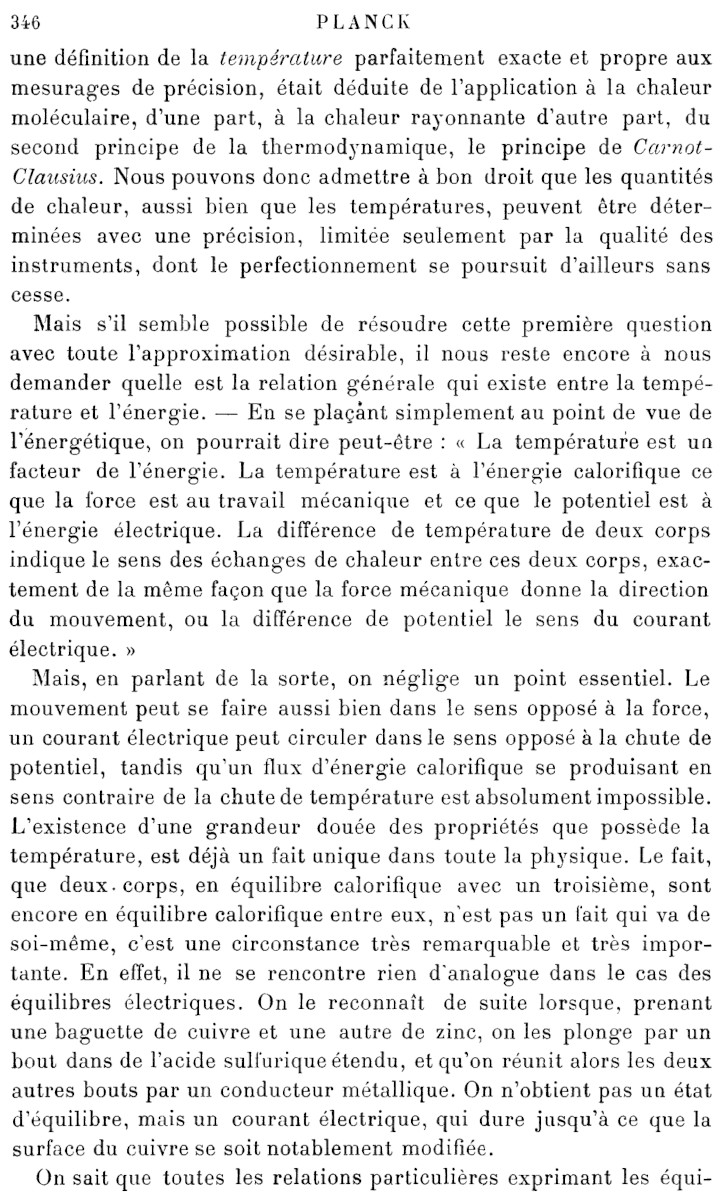}
\end{figure}
\begin{figure}[hbt]
\centering
\includegraphics[width=0.77\linewidth]{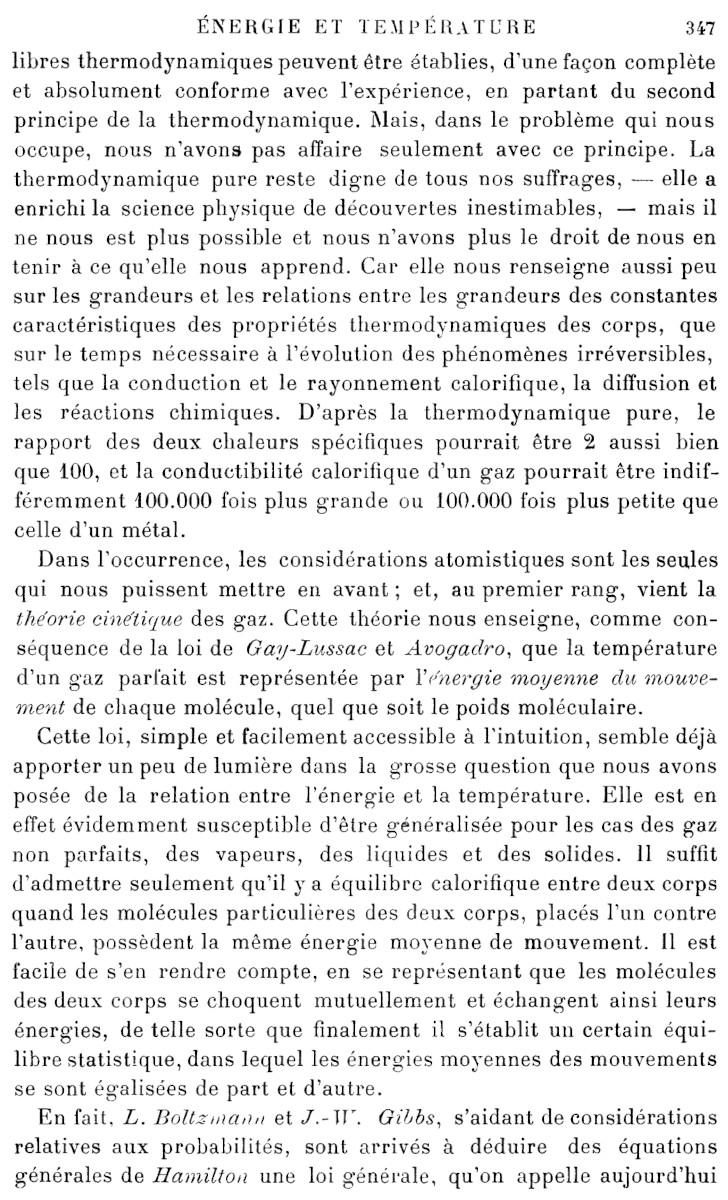}
\end{figure}
\begin{figure}[hbt]
\centering
\includegraphics[width=0.77\linewidth]{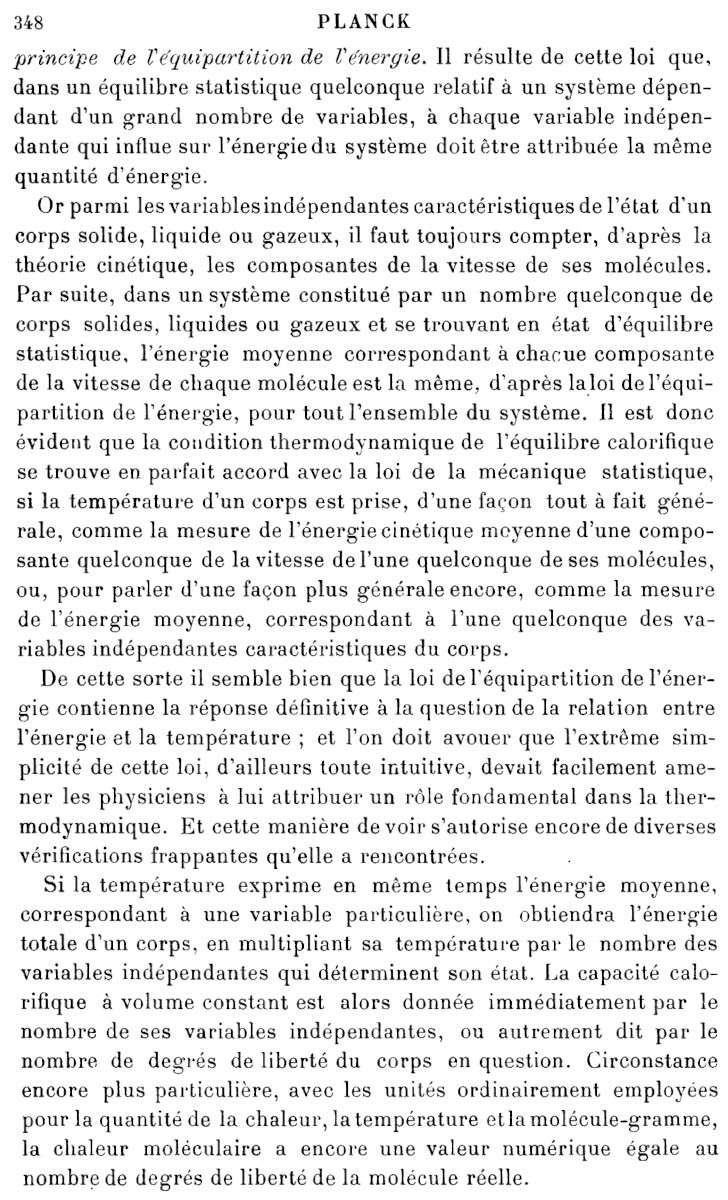}
\end{figure}
\begin{figure}[hbt]
\centering
\includegraphics[width=0.77\linewidth]{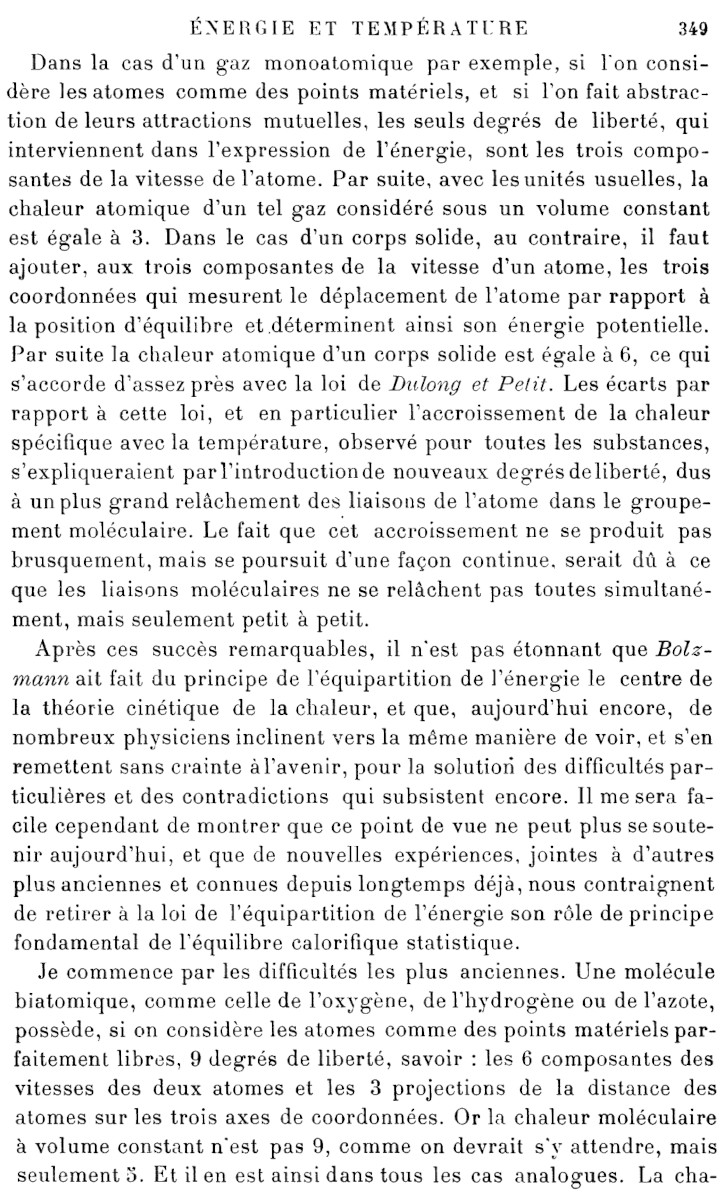}
\end{figure}
\begin{figure}[hbt]
\centering
\includegraphics[width=0.77\linewidth]{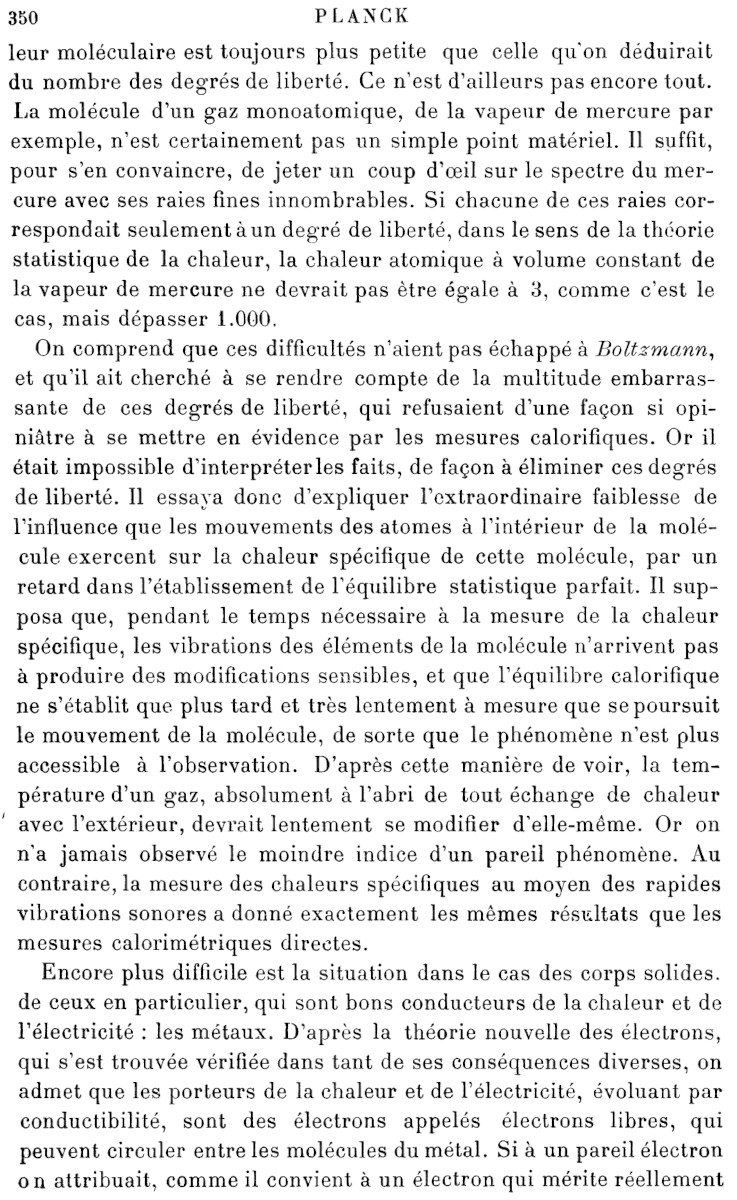}
\end{figure}
\begin{figure}[hbt]
\centering
\includegraphics[width=0.77\linewidth]{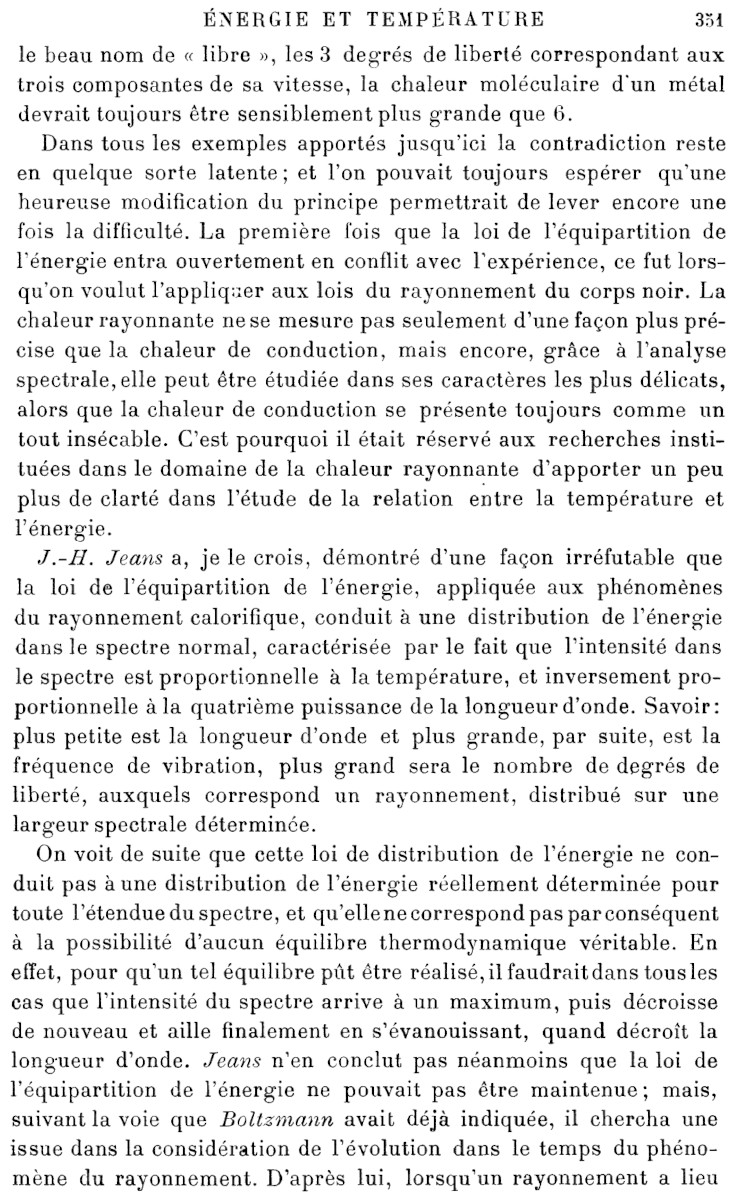}
\end{figure}
\begin{figure}[hbt]
\centering
\includegraphics[width=0.77\linewidth]{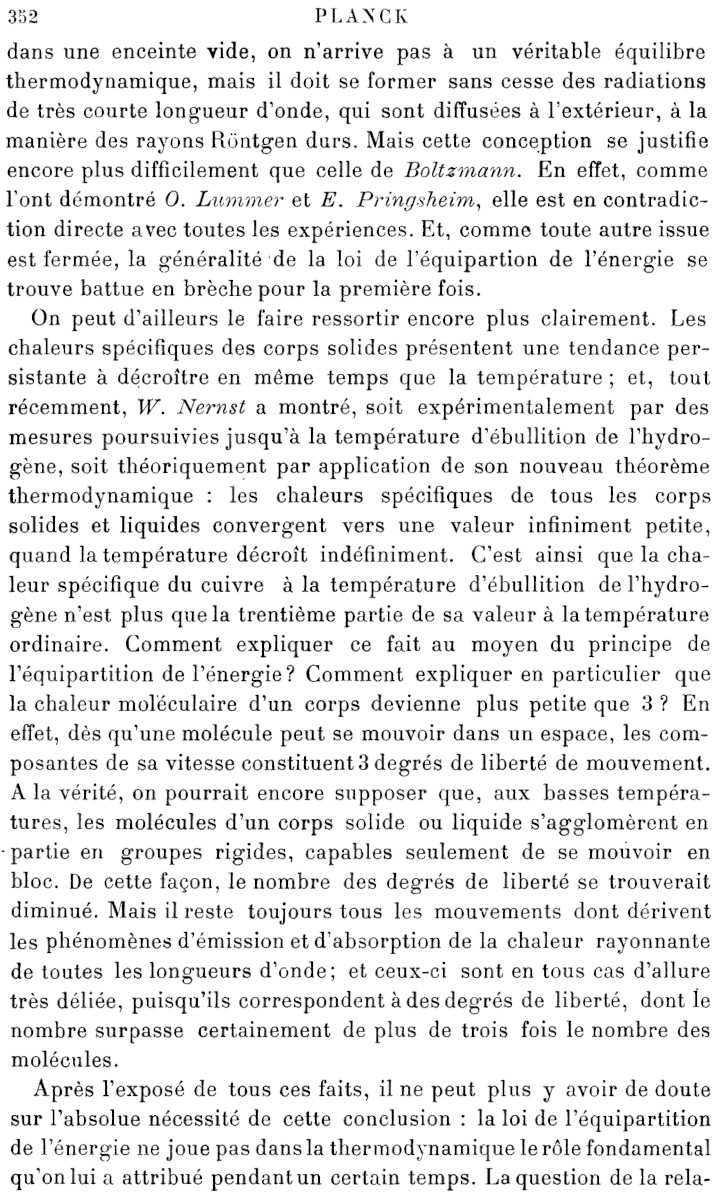}
\end{figure}
\begin{figure}[hbt]
\centering
\includegraphics[width=0.77\linewidth]{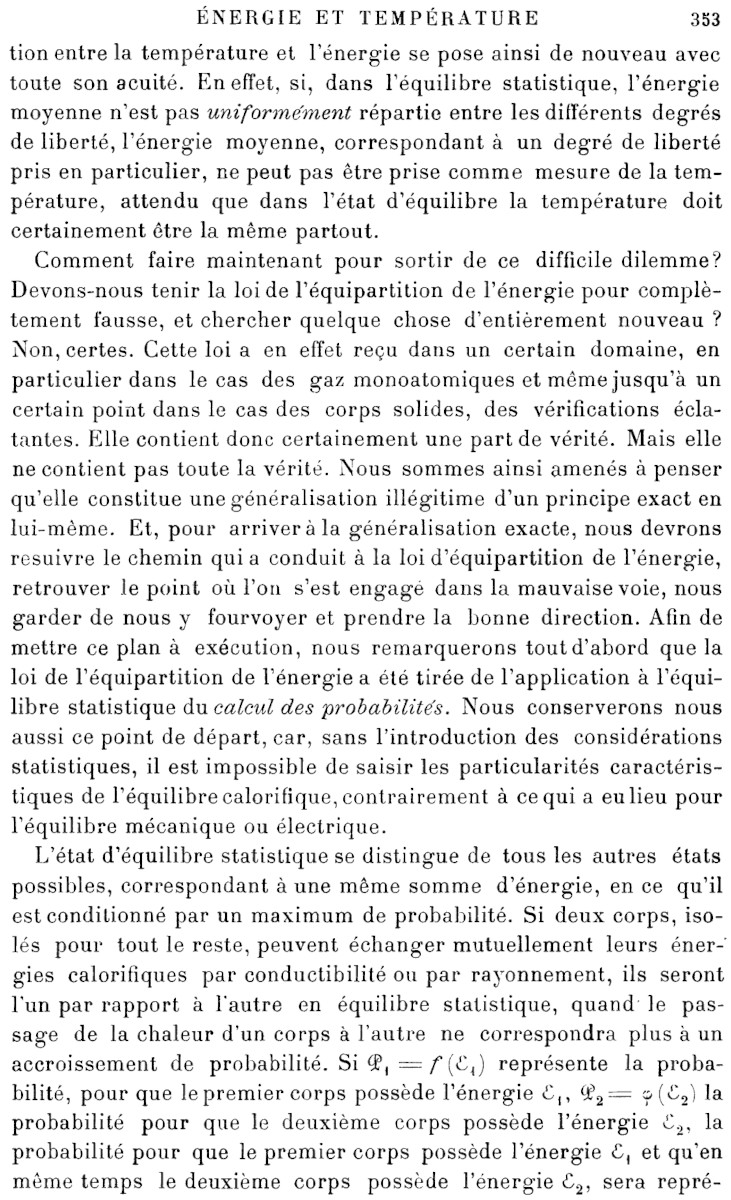}
\end{figure}
\begin{figure}[hbt]
\centering
\includegraphics[width=0.77\linewidth]{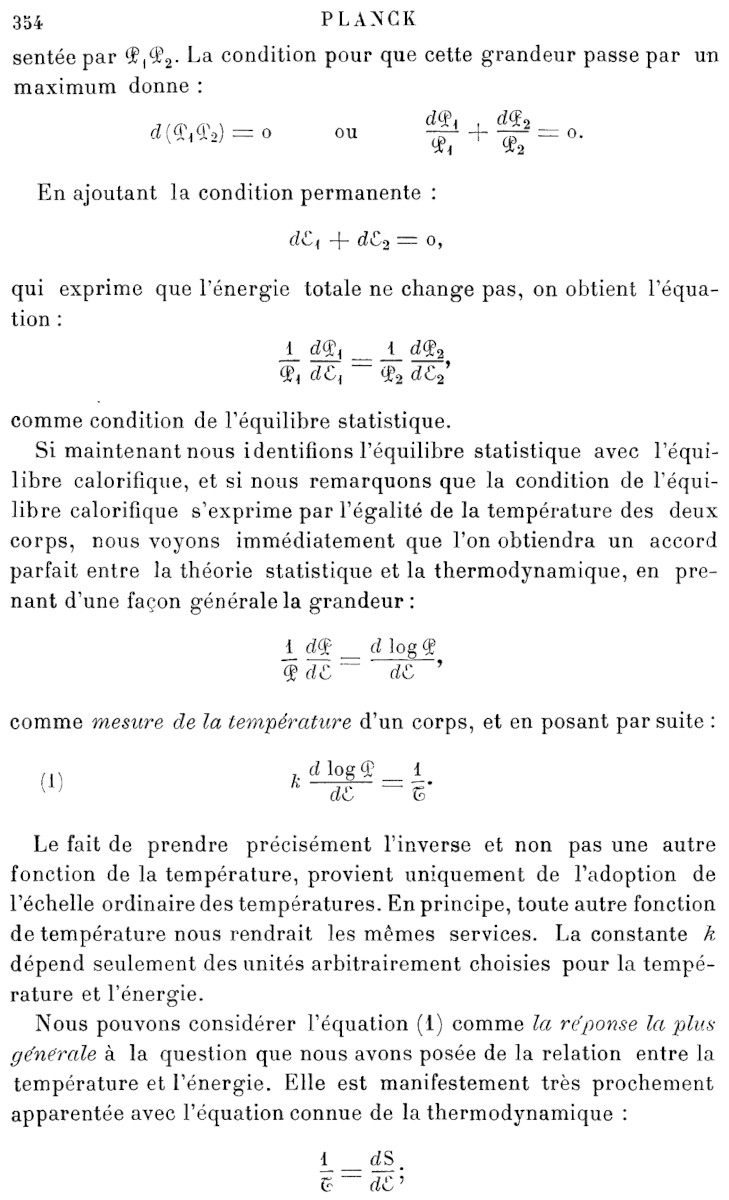}
\end{figure}
\begin{figure}[hbt]
\centering
\includegraphics[width=0.77\linewidth]{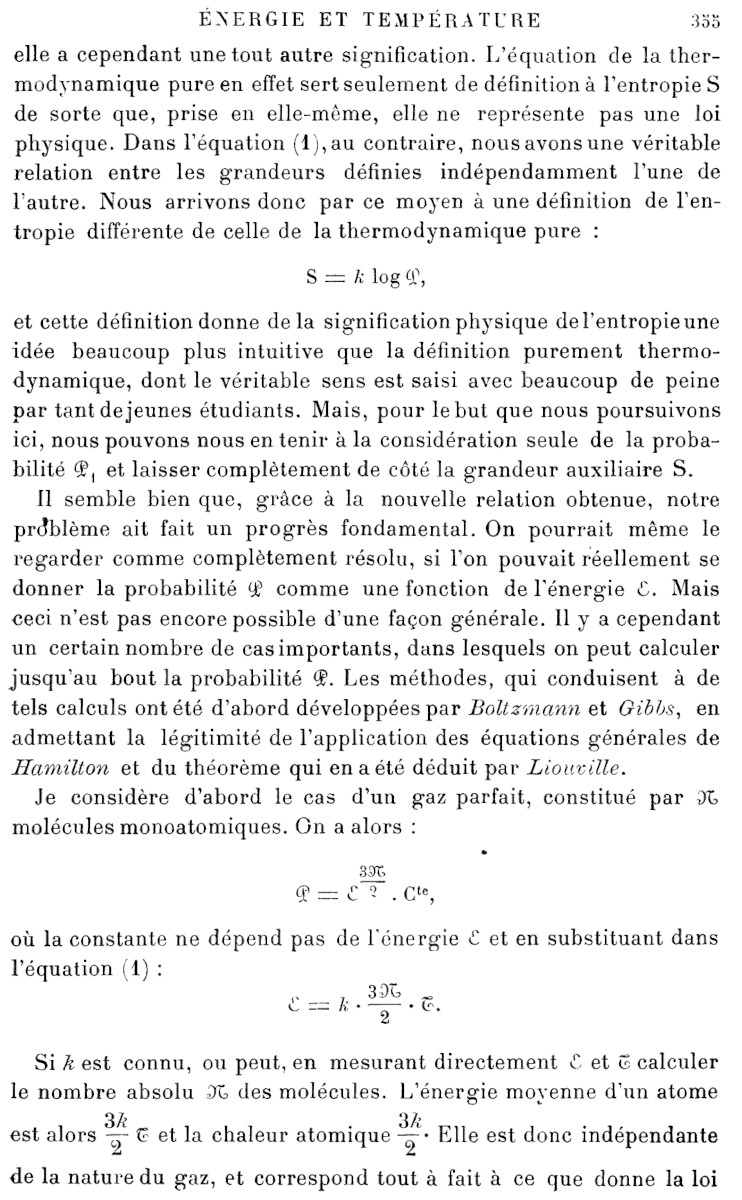}
\end{figure}
\begin{figure}[hbt]
\centering
\includegraphics[width=0.77\linewidth]{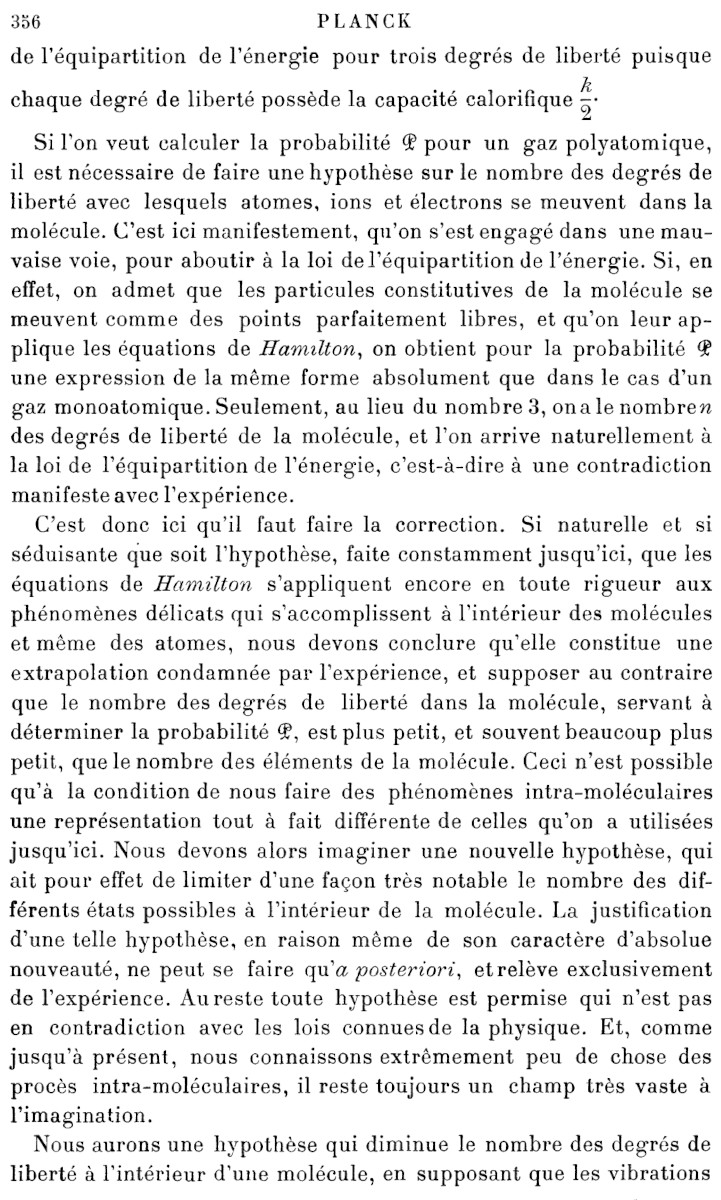}
\end{figure}
\begin{figure}[hbt]
\centering
\includegraphics[width=0.77\linewidth]{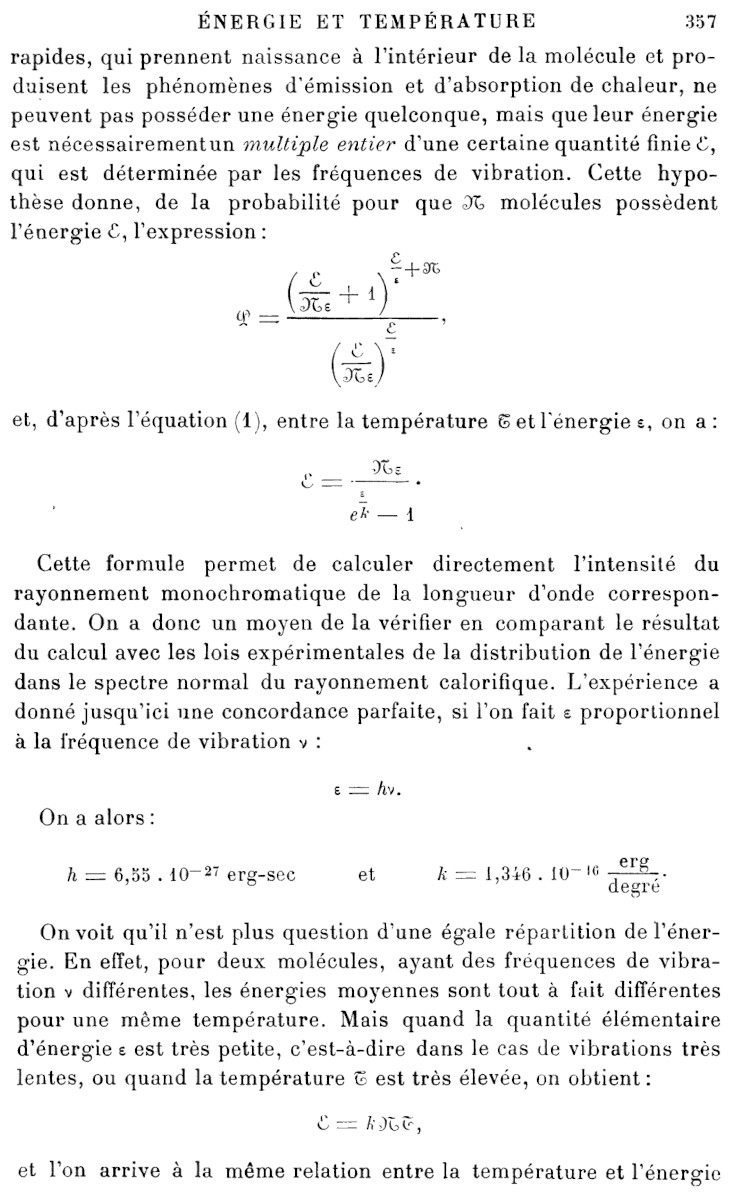}
\end{figure}
\begin{figure}[hbt]
\centering
\includegraphics[width=0.77\linewidth]{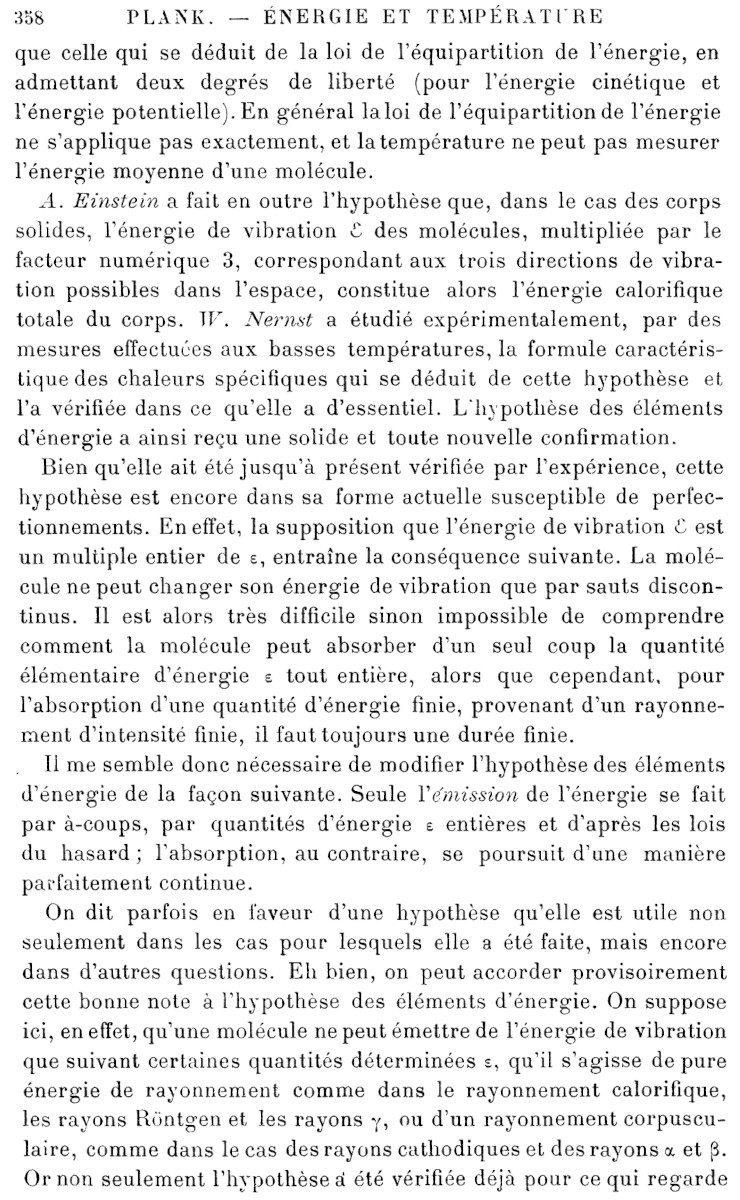}
\end{figure}
\begin{figure}[hbt]
\centering
\includegraphics[width=0.77\linewidth]{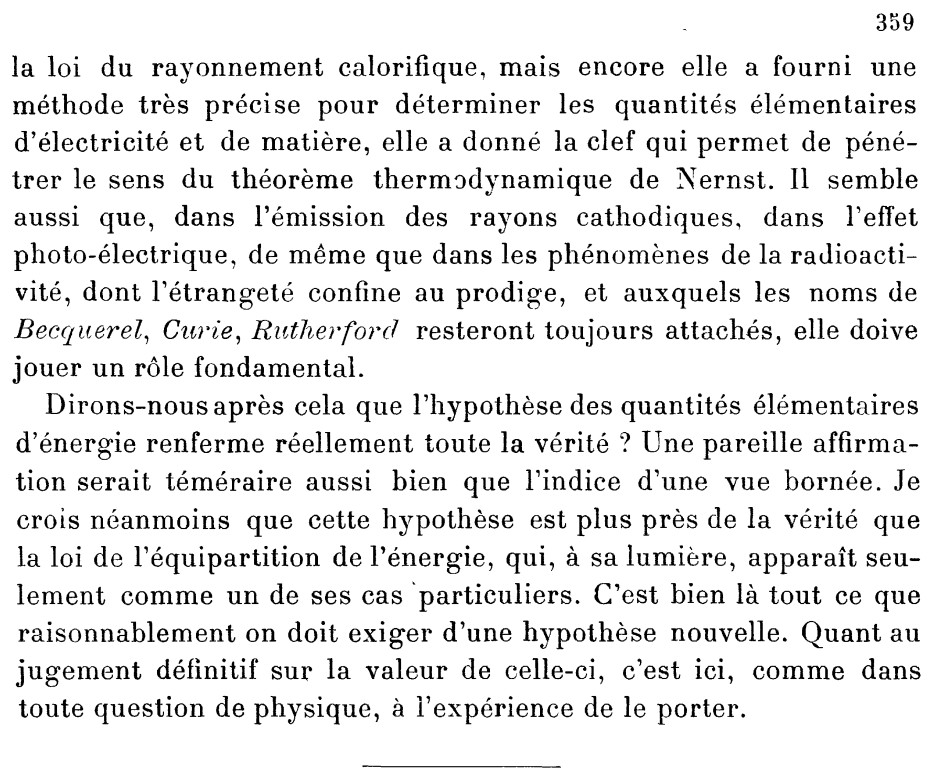}
\end{figure}

\clearpage \newpage
\section{\underline{\color{blue} The German Paper of Planck (1911)}}
\label{Section-German-PDF-1911}
\vspace*{0mm}

\begin{figure}[hbt]
\centering
\includegraphics[width=0.77\linewidth]{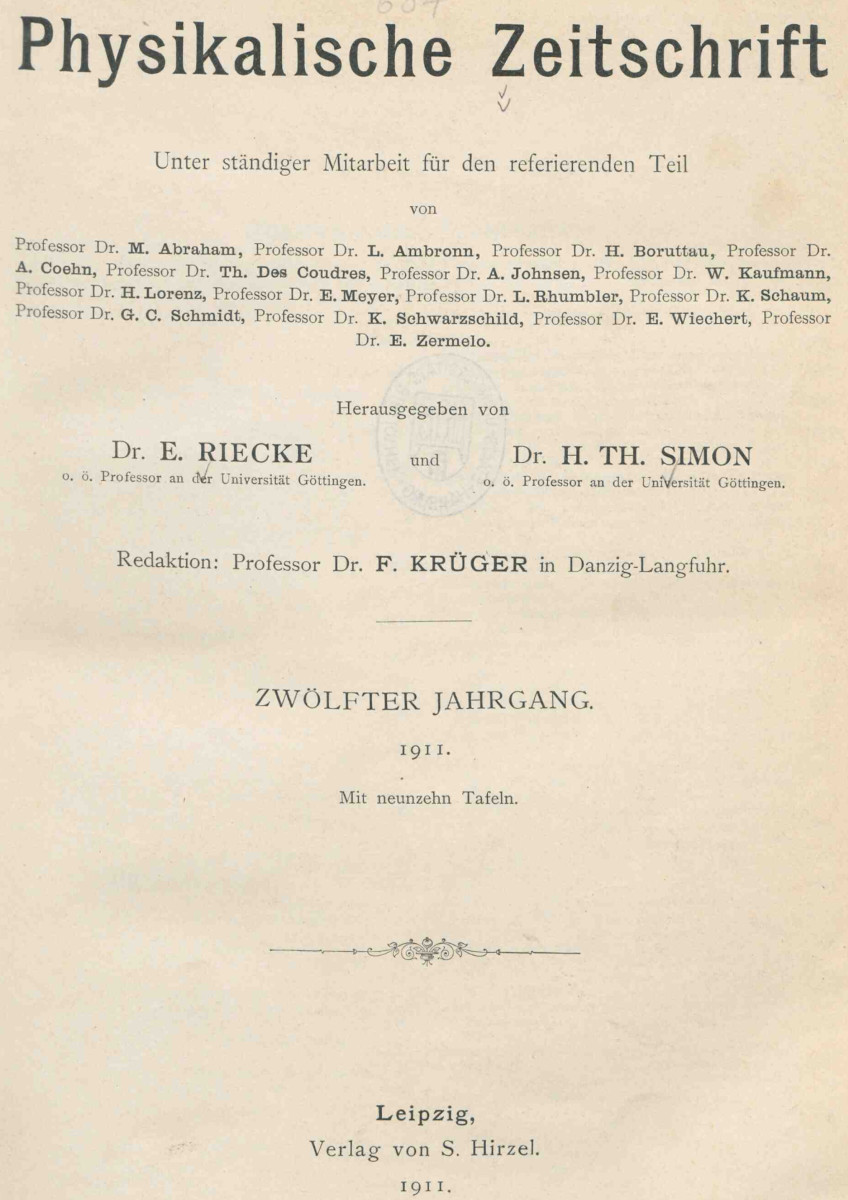}
\end{figure}
\begin{figure}[hbt]
\centering
\includegraphics[width=0.77\linewidth]{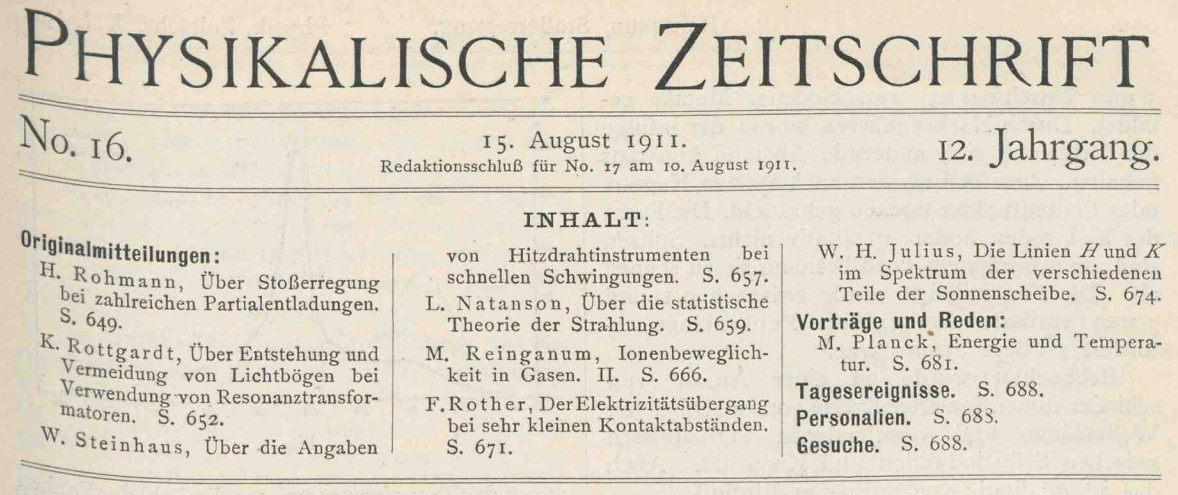}
\includegraphics[width=0.77\linewidth]{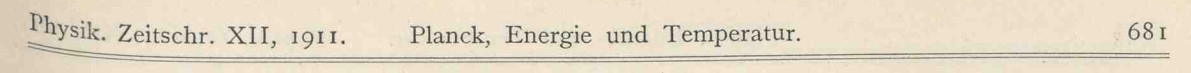}
\includegraphics[width=0.77\linewidth]{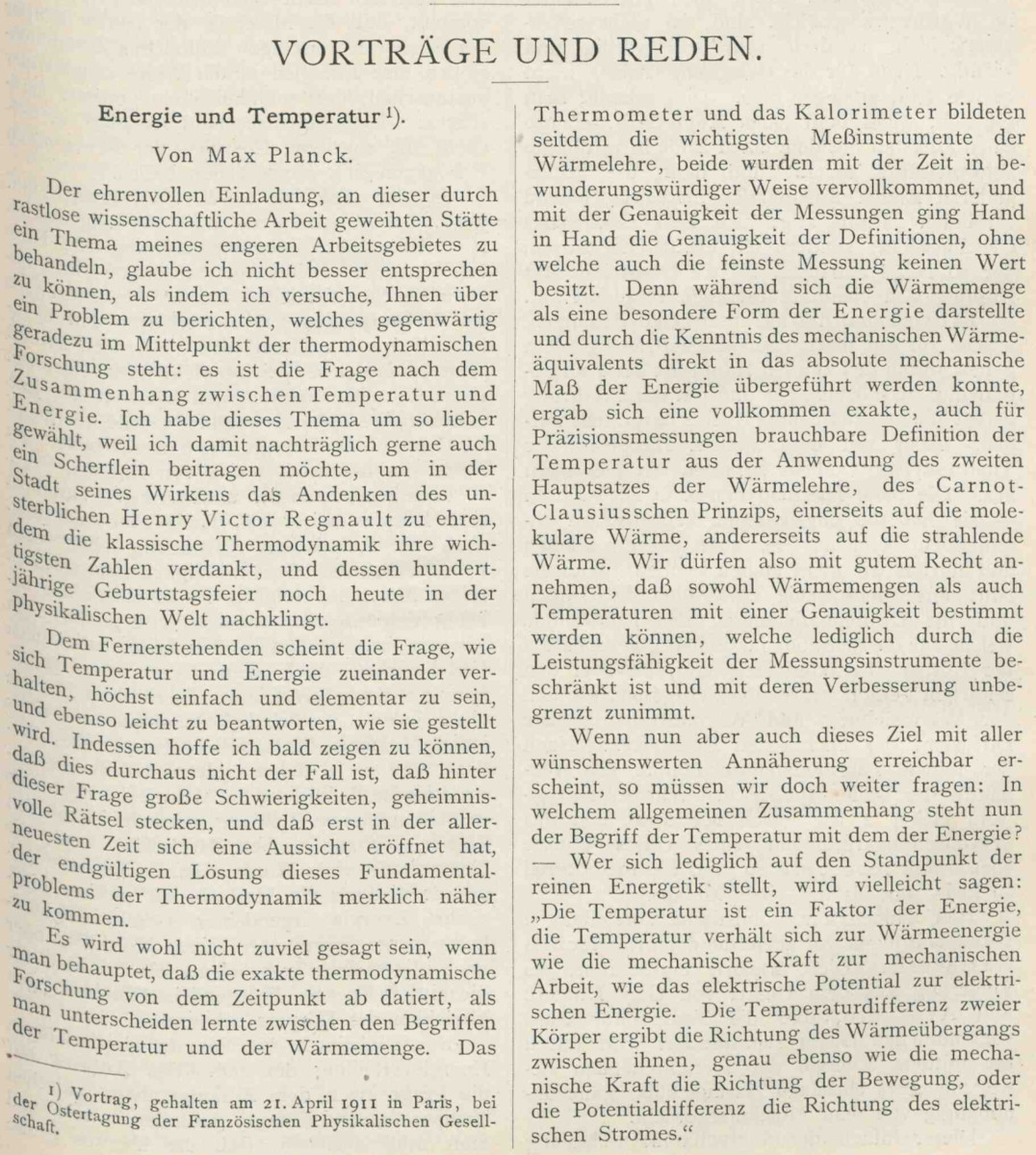}
\end{figure}
\begin{figure}[hbt]
\centering
\includegraphics[width=0.77\linewidth]{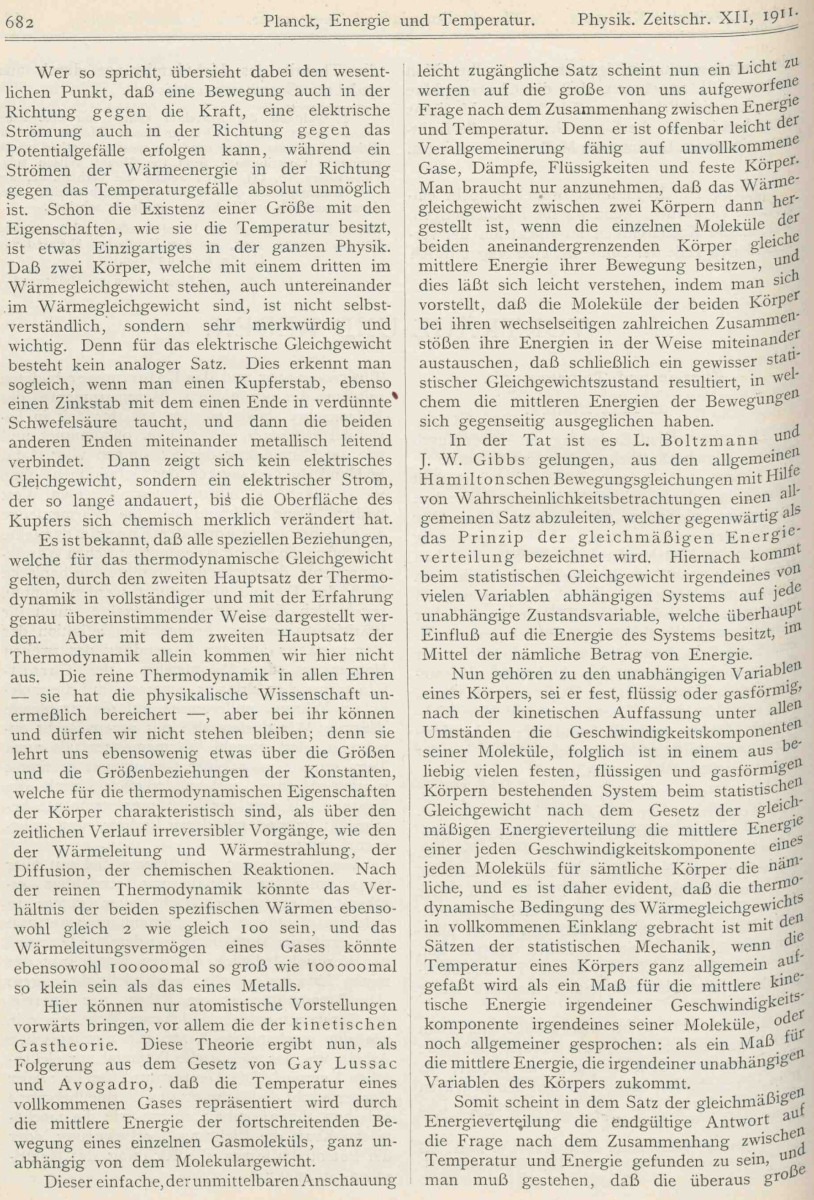}
\end{figure}
\begin{figure}[hbt]
\centering
\includegraphics[width=0.77\linewidth]{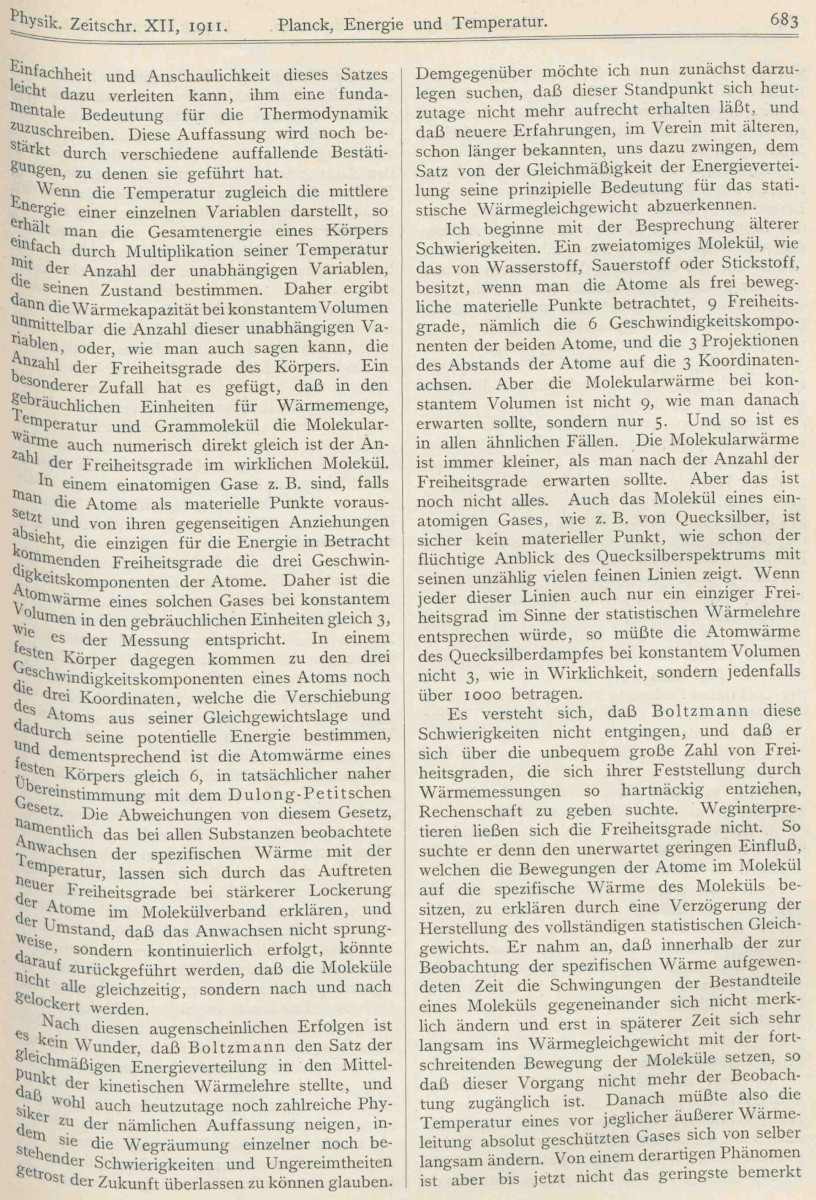}
\end{figure}
\begin{figure}[hbt]
\centering
\includegraphics[width=0.77\linewidth]{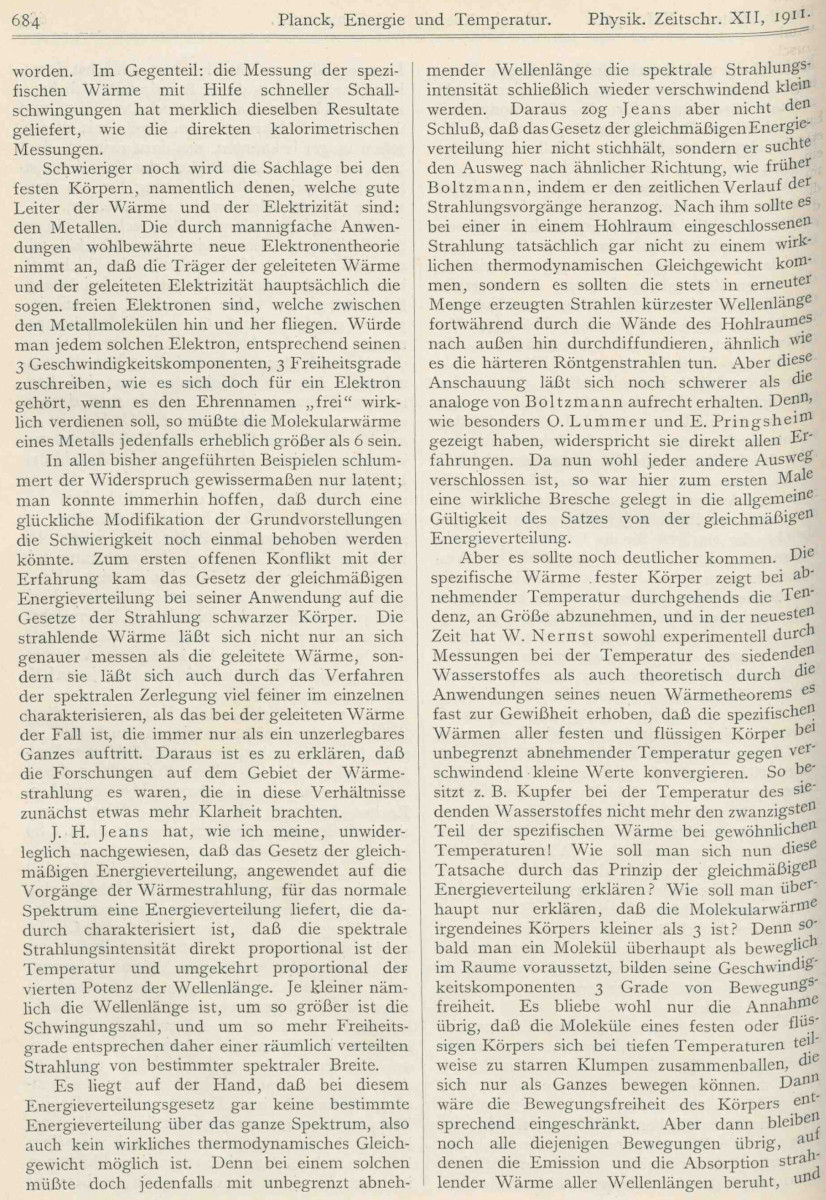}
\end{figure}
\begin{figure}[hbt]
\centering
\includegraphics[width=0.77\linewidth]{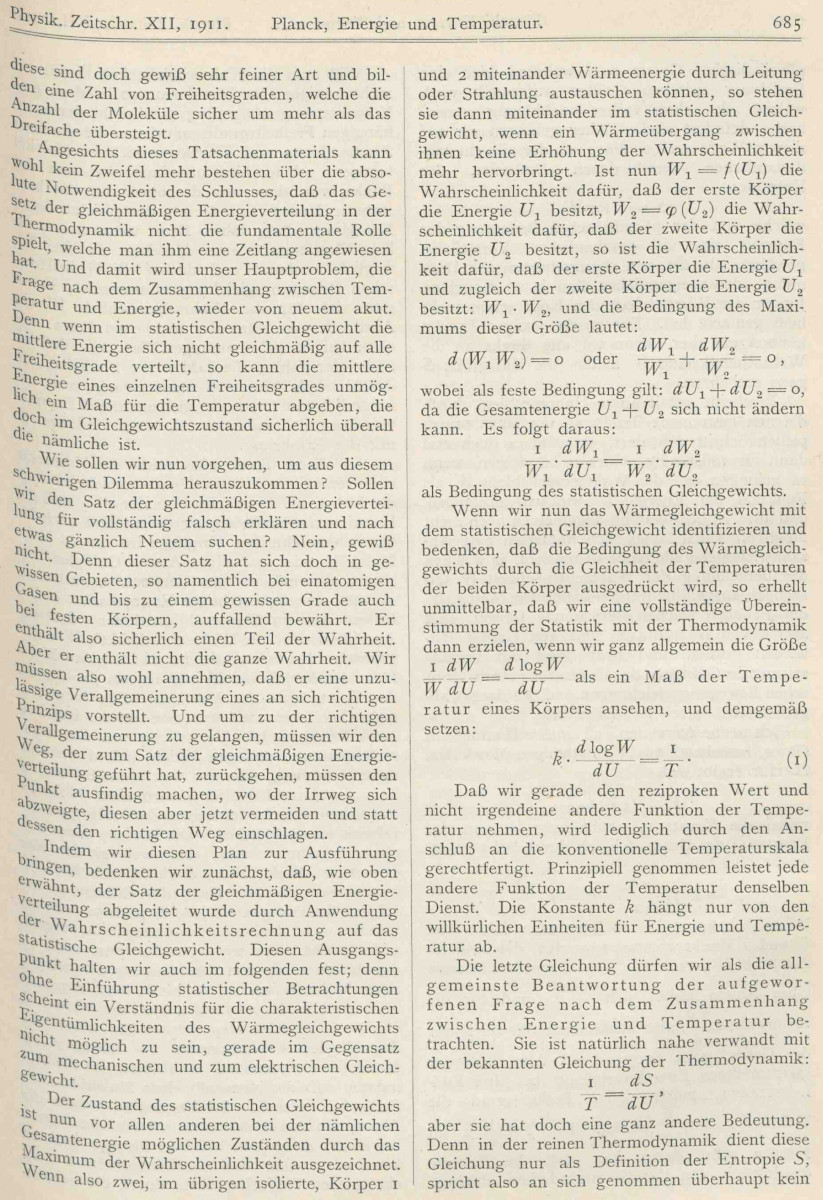}
\end{figure}
\begin{figure}[hbt]
\centering
\includegraphics[width=0.77\linewidth]{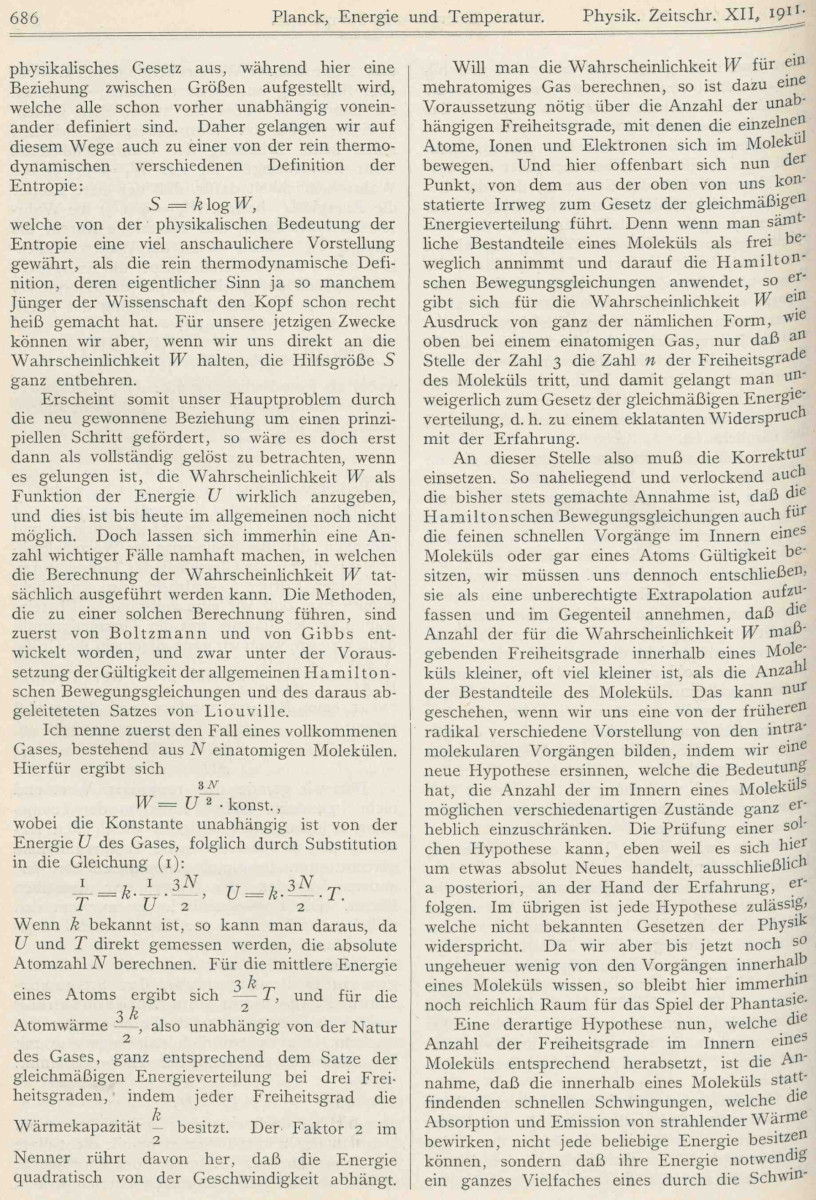}
\end{figure}
\begin{figure}[hbt]
\centering
\includegraphics[width=0.77\linewidth]{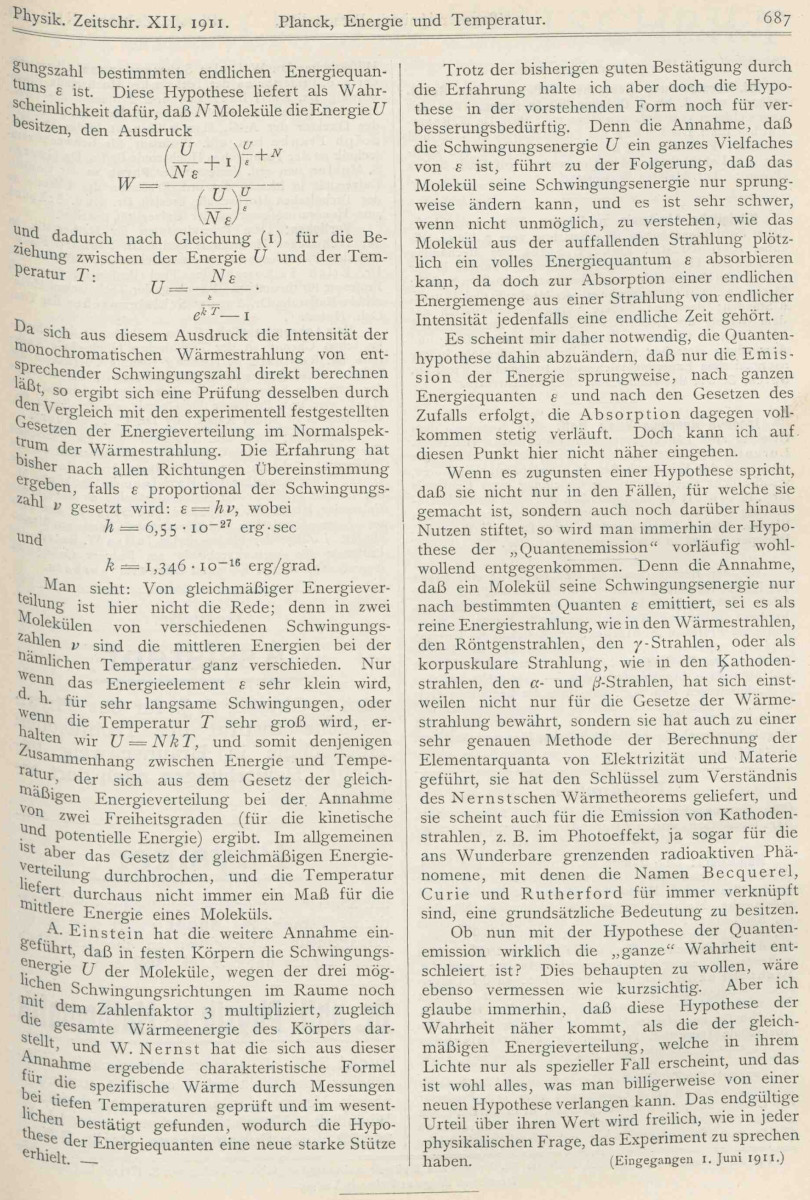}
\end{figure}

\end{document}